\documentclass[twocolumn,pra,superscriptaddress,showpacs,showkeys,amsmath,amssymb]{revtex4-1}
\usepackage{graphicx}

\usepackage{color}

\newcommand{\rf}[1]{(\ref{#1})}
\newcommand{\Eta}{\mathop{\rm H}\nolimits}
\newcommand{\sech}{\mathop{\rm sech}\nolimits}
\newcommand{\sgn}{\mathop{\rm sgn}\nolimits}

\newcommand{\real}{\mathop{\rm Re}\nolimits}
\newcommand{\erf}{\mathop{\rm erf}\nolimits}
\newcommand{\Prob}{\mathop{\rm Prob}\nolimits}
\newcommand{\oh}{\frac{1}{2}}
\newcommand{\expe}{\mathrm{e}}
\newcommand{\difd}{\mathrm{d}}

\newcommand{\xp}{x^{\prime}}

\newcommand{\Ralph}{R_{\alpha}}

\newcommand{\lcor}{L_{\mathrm{c}}}

\newcommand{\lswint}{L_{\mathrm{int}}}

\newcommand{\lswtrans}{L_{\mathrm{tra}}}
\newcommand{\lswfirst}{L_{\mathrm{lin},0}}

\newcommand{\Xdepth}{X_{\mathrm{dep}}}
\newcommand{\Xtra}{X_{\mathrm{tra}}}
\newcommand{\Xint}{X_{\mathrm{int}}}
\newcommand{\Xstay}{X_{\mathrm{fluc}}}
\newcommand{\Xlinfirst}{X_{\mathrm{lin},0}}
\newcommand{\Xret}{X_{\mathrm{ret}}}

\newcommand{\pXtra}{p_{\Xtra}}
\newcommand{\pXint}{p_{\Xint}}
\newcommand{\pXret}{p_{\Xret}}
\newcommand{\plinfirst}{p_{\Xlinfirst}}
\newcommand{\La}{L_{a}}
\newcommand{\Lb}{L_{b}}
\newcommand{\Lab}{L_{\mathrm{fluc}}}
\newcommand{\Lfluc}{\Lab}

\newcommand{\st}{\mathcal T}

\newcommand{\etainit}{\eta_0}
\newcommand{\tmax}{t_{\max}}
\newcommand{\phip}{\phi^{\prime}}
\newcommand{\lfluc}{L_{\mathrm{fluc}}}

\begin{document}

\title{Stochastic Pulse Switching in a Degenerate Resonant Optical Medium}
\author{Ethan P. Atkins}\affiliation{Department of Mathematics, University of California, Berkeley, 970 Evans Hall \#3840, Berkeley, CA  94720-3840}
\author{Peter R. Kramer}
\author{Gregor Kova\v{c}i\v{c}}\email{kovacg@rpi.edu}\affiliation{Mathematical Sciences Department, Rensselaer Polytechnic Institute, 110 8th Street, Troy, NY 12180}
\author{Ildar R. Gabitov}\affiliation{Department of Mathematics, The University of Arizona,
617 N. Santa Rita Ave., P.O. Box 210089,  Tucson, AZ 85721-0089}\affiliation{Department of Mathematics, Southern Methodist University, 3200 Dyer Street,  P.O. Box 750156,  Dallas TX 75275-0156}

\begin{abstract}
Using the idealized integrable Maxwell-Bloch model, we describe random optical-pulse polarization switching along an active optical medium in the $\Lambda$-configuration with disordered occupation numbers of its lower energy sub-level pair.  The description combines complete integrability and stochastic dynamics.  For the single-soliton pulse, we derive the statistics of the electric-field polarization ellipse at a given point along the medium in closed form.     If the average initial population difference of the two lower sub-levels vanishes, we show that the pulse polarization will switch intermittently between the two circular polarizations as it travels along the medium.   If this difference does not vanish, the pulse will eventually forever remain in the circular polarization determined by which sub-level is more occupied on average.   We also derive the exact expressions for the statistics of the polarization-switching dynamics, such as the probability distribution of the distance between two consecutive switches and the percentage of the distance along the medium the pulse spends in the elliptical polarization of a given orientation in the case of vanishing average initial population difference.  We find that the latter distribution is given in terms of the well-known arcsine law. 
\end{abstract}



\maketitle

\section{Introduction}
Resonant interaction of light with active optical media has given rise to one of the most fruitful areas of applied physics in having provided the foundation
of numerous important physical effects over the past six decades~\cite{ISI:A1960ZQ06900019,kurnit64,mccall67,mccall69,%
PhysRevA.5.1634,PhysRev.179.294,PhysRevLett.29.1211,%
PhysRevLett.30.309,PhysRevLett.36.1035,PhysRevLett.39.547,%
PhysRevLett.57.2804,%
PhysRevLett.66.2593,harris:36,%
hau99,Slowlight99}, and served as one of the basic mechanisms used in laser operation and optical amplifiers~\cite{shimoda86,allen87,butcher90,Boyd92,Newell92}.   While a fully quantum description of this interaction has also been developed~\cite{ruyu84,allen87},  probably its most revealing description has been furnished by the semiclassical model provided by the Maxwell-Bloch system of equations~\cite{Feynman57,davis63,Jaynes63,risken:4662,allen87}.   This model has helped to uncover the fundamentals of the resonant interaction between an electromagnetic field and a system of active atoms in the regime in which the field can be described classically and the medium by quantum mechanics, and in which a great number of the relevant experiments have been carried out~\cite{ISI:A1960ZQ06900019,kurnit64,mccall67,mccall69,%
PhysRevA.5.1634,PhysRev.179.294,PhysRevLett.29.1211,%
PhysRevLett.30.309,PhysRevLett.36.1035,PhysRevLett.39.547,%
PhysRevLett.57.2804,%
PhysRevLett.66.2593,harris:36,%
hau99,Slowlight99}.   In fact, many physical effects observed in these experiments,  from the photon echo~\cite{kurnit64} and self-induced transparency~\cite{mccall67,mccall69} to the chaotic laser dynamics~\cite{PhysRevLett.57.2804},  have been explained using the Maxwell-Bloch equations in the idealized two-level approximation, in which the light is assumed to be monochromatic and to interact resonantly with only two active atomic levels in the  optical medium~\cite{davis63,Jaynes63,risken:4662,kaup77,ISI:A1984TY33700007,allen87,Milonni05}.

In the simplest case of the two-level approximation, when the pulses interacting with the medium are much shorter than the relaxation times of the medium, the Maxwell-Bloch system describing this interaction is completely integrable~\cite{ablowitz74}.     This feature was used to theoretically explain
three important phenomena: self-induced transparency, superfluorescence, and quantum amplification.
The McCall-Hahn phenomenon of self-induced transparency~\cite{mccall67,mccall69,PhysRevA.5.1634}---a medium whose atoms are initially in the ground state becoming transparent to optical pulses with the resonant carrier frequency---was first analyzed  using complete integrability in~\cite{ablowitz74}, after many exact solutions hinting at possible integrability had been found in~\cite{lamb71,lamb74}.  For  superfluorescence~\cite{PhysRevLett.30.309,PhysRevLett.36.1035,PhysRevLett.39.547}---the generation of optical pulses from the random fluctuations of the initial medium polarization  in an excited medium---the linear stage was addressed in~\cite{PhysRevA.23.1322}, where the statistics of the delay time between the pumping of the medium and the pulse maximum were derived in terms of the statistics of the polarization fluctuations, and shown to be Gaussian.
The fully nonlinear problem was subsequently addressed using its integrable structure in~\cite{gabitov83,gabitov84,gabitov85}, whose main result was the shape of the superfluorescence pulse and its relation to the delay time.   The fully nonlinear description of a quantum amplifier---incident-pulse amplification in an excited medium---was addressed in~\cite{manakov82,manakov86}.

An approximate description of the medium as having more than two levels, or degenerate levels, interacting with the  light pulses propagating through it, is more physically realistic than the idealized two-level approximation.   Such a description is desirable, for example, because effects such as self-induced transparency have also been measured for transitions between degenerate levels~\cite{PhysRevLett.27.287}.  An important special active medium with a doubly-degenerate ground level and an excited level as its working levels is referred to as the $\Lambda$-configuration medium, so named because of the shape of the corresponding quantum transition diagram.   The two types of atomic transitions in such a medium are stimulated by and emit light of two opposite circular polarizations~\cite{konopnicki81,konopnicki81a}.  The complete integrability of the Maxwell-Bloch systems describing light pulses interacting with this type of a medium was discovered in~\cite{bolshov83,basharov84,maimistov84,Chernyak1985434}, and self-induced transparency was described.   Superfluorescence and amplification of incident pulses via the transfer of energy from the initially excited medium to the pulse in $\Lambda$-configuration media were studied in~\cite{gabitov88} and~\cite{GabitovManakov83}, respectively.

One feature distinguishing the $\Lambda$-configuration description from the simpler two-level description is its ability to capture the polarization of the propagating pulses, and thus polarization switching which depends on the initial population of the two degenerate lower levels~\cite{Maimistov85,Byrne03}.   Another distinguishing feature of the corresponding one-soliton solution is that it is a soliton only in the sense of being a potential in the direct scattering problem of the corresponding Lax operator that gives rise to a single-eigenvalue spectrum, but is not a solitary traveling wave.   In fact, even in the simplest case of constant initial lower-level populations, its shape is only asymptotically stationary.  It has complex internal dynamics through which both its shape and velocity can change in space and time, and thus can reflect the light polarization switching.

In this paper, using the corresponding integrable Maxwell-Bloch system, we describe random polarization switching of pulses passing through a $\Lambda$-configuration medium induced by a disordered initial population of the degenerate lower sub-levels.
The dependence of the properties of an emerging light pulse on both integrability as well as randomness in the initial conditions appears already for the simpler ideal two-level optical medium  through the phenomenon of superfluorescence~\cite{gabitov83,gabitov84,gabitov85}.  Randomness in a two-level medium however appears to play a negligible role in self-induced transparency.    Richer interactions between the integrable dynamics and random initial data emerge in a $\Lambda$ configuration medium, as the flexibility of populating the degenerate lower sub-levels makes it possible for self-induced transparency to take place in the presence of structural disorder  arising from spatial fluctuations in these populations.  
Such disorder often results during the initial preparation of the resonant medium, and subsequently induces random polarization switching of light pulses propagating through this medium.  We will compute and analyze several statistical properties  of this nonlinear random polarization switching using exact results obtained with the inverse-scattering-transform technique for the $\Lambda$-configuration Maxwell-Bloch equations.

While, in reality, a pulse propagating through a resonant active medium encounters several sources of losses that make it decay on a number of relaxation scales, we have chosen its idealized lossless integrable Maxwell-Bloch description for two reasons.    The first is that we aim to understand the fundamental features of the polarization switching process in the course of this propagation, in particular how the pulse and medium parameters affect its statistical properties.  The second is that, at the current development level of the experimental instrumentation, the situation in which the pulse width is much shorter than the relaxation times is achievable, so that our model should be realistic from the  viewpoint of experiments.  Therefore, we here consider the idealized integrable model describing pulse interaction with a degenerate active optical medium in the $\Lambda$ configuration with structural disorder introduced by an inhomogeneous distribution of the degenerate lower sub-level population in the medium.

We present the polarization statistics for the one-soliton solution, both because we can obtain them explicitly and because they yield a particularly transparent description of the switching phenomenon.   In our treatment, we use the classical polarization ellipse representation~\cite{born,Jackson75}, which has the advantage of being independent of time for the one-soliton solution; in other words, for the single-soliton pulse, the two angles determining the shape of the polarization ellipse only depend on the location along the medium sample.  We address the statistics of the pulse travel-time to a given location along the medium sample, the shape statistics of the polarization-ellipse  at that location, and also the statistics of the switching between the left- and right-circular polarizations that a soliton pulse experiences while traveling along the sample.

We explore the qualitatively different statistical dynamical regimes that emerge for the cases when the initial degenerate-lower-level populations have (approximately) equal  or unequal mean.  
  In particular, we find that, when the lower levels are equally populated on average, the polarization lingers close to one of the two circular polarizations for long distances but can forever switch intermittently between the two with probability one.  On the other hand,  when   the initial degenerate-lower-level populations along the optical medium have distinct averages, the polarization after a few possible initial switches asymptotically approaches the circular polarization corresponding to the transition between the on-average initially less populated lower sub-level and the excited level, and no further switching occurs with probability one.

The remainder of the paper is organized as follows.   In Sec.~\ref{sec:background} we present the relevant background of the problem.   In particular, in Sec.~\ref{sec:polarization_dyn}, we review the polarization ellipse representation of polarized light, in Sec.~\ref{sec:mbes}, we review the Maxell-Bloch equations that describe resonant interaction of pulses with a $\Lambda$-configuration degenerate active optical medium, and in Sec.~\ref{sec:cauchy} we we review the inverse-scattering-transform method and soliton solution used in the description of light-polarization dynamics.   In Sec.~\ref{population:disorder} we discuss the soliton statistics when the initial medium population is disordered in space, with the approximate white-noise description presented in Sec.~\ref{sec:wienercoarse}, and the statistics of the soliton travel time to a given point and the two angles determining the shape of the polarization ellipse at that point discussed in Sec.~\ref{sec:wienerstats}.   The statistical description of the polarization switching dynamics is given in Sec.~\ref{sec:dynpol}, and concluding remarks are presented in Sec.~\ref{sec:conclusions}.  Appendix~\ref{sec:lcor} further elucidates the appearance and role of the correlation length of the initial lower-level population difference along the optical medium.   Appendix~\ref{sec:conv} contains a detailed calculation of two polarization-switching length statistics.

\section{Background and Problem Formulation\label{sec:background}}

In this section, we describe the problem at hand by briefly reviewing the polarization ellipse description of polarized light, the three-level Maxwell-Bloch equations that describe the propagation of monochromatic, elliptically-polarized light through a $\Lambda$-configuration active optical medium, and the soliton solution whose random polarization switching dynamics we will study in the rest of the paper.

\subsection{Optical Pulse Polarization}
\label{sec:polarization_dyn}

The electric field polarization is among the light characteristics most sensitive to changes in the properties of the optical medium.    This makes it a good potential target for experimental investigation of the stochastic behavior of the light pulses predicted in this paper.   Therefore, in this section, we present a brief discussion of its main properties of importance to our subsequent discussion.

\begin{figure}[h]\begin{center}
\includegraphics[scale=.32]{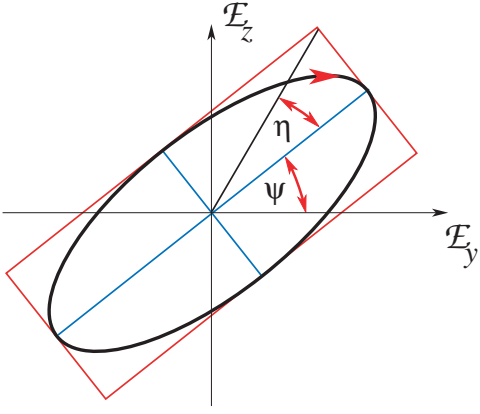}
\caption{\label{fig:ellipse}Polarization ellipse in the plane
perpendicular to the direction of light propagation.}\end{center}
\end{figure}

Two well-established descriptions of light polarization are given in terms of the polarization ellipse and Poincar\'e sphere~\cite{born,Jackson75}.
Here, we review the basic concepts of the polarization ellipse description, which we will use in the rest of the paper. (See also~\cite{gabitov88}.)
To this end, we consider a plane electromagnetic wave with
frequency $\omega$ and wave number
$k$, propagating in the positive
$x$ direction in the $(x,y,z,t)$-laboratory coordinate frame, which has the form
\begin{eqnarray}&& \vec{\mathcal E} (x,t) = \real \left\{\vec E e^{i(kx-\omega t)}\right\}\nonumber\\ &&=\real
\left\{\Bigl[E_y \vec{\mathbf e}_y + E_z \vec{\mathbf e}_z\Bigr] e^{i(kx-\omega t)}
\right\}.
\label{evec} \end{eqnarray}
Here, $\real $ denotes the real part of a complex number, $E_y$ and $E_z$ are the two components of the complex wave amplitude, $\vec E$, and
$\vec{\mathbf e}_y$ and $\vec{\mathbf e}_z$ are the unit vectors in the $(y,z)$-plane,
perpendicular to the propagation direction of the wave. Defining the
circular-polarization basis vectors $\vec {\mathbf e}_+$ and $\vec
{\mathbf e}_-$ as
\begin{equation}\label{circbasis}
\vec {\mathbf e}_+ =\frac{1}{\sqrt 2} \left(\vec{\mathbf e}_y + i
\vec{\mathbf e}_z\right),\qquad \vec {\mathbf e}_- =\frac{1}{\sqrt 2} \left(\vec{\mathbf e}_y - i
\vec{\mathbf e}_z\right),\end{equation}
we rewrite the complex wave amplitude $\vec E$ in the
circular-polarization components as
\begin{eqnarray}
&&\vec E = E_y \vec {\mathbf e}_y + E_z \vec {\mathbf e}_z=
E_+\,\vec {\mathbf e}_+ + E_-\,
\vec {\mathbf e}_- \nonumber \\ &&   = |E_+| e^{i\phi}\, \vec {\mathbf e}_+ +
|E_-|  e^{i(\phi+ 2\psi)}\,\vec {\mathbf e}_- \nonumber\\ &&= \left(|E_+| e^{-i\psi}\, \vec {\mathbf
e}_+ + |E_-|  e^{i
\psi} \,\vec {\mathbf e}_-\right) e^{i(\phi+\psi)},\label{polarized}
\end{eqnarray}
where $\phi$ and $\phi+2\psi$ are the phases of the $\vec {\mathbf e}_+$ and $\vec {\mathbf e}_-$ electric field components, respectively.
This yields the expression for the electric field
\begin{eqnarray}
&&\vec {\mathcal E} (x,t)= \frac{|E_+| + |E_-|}{\sqrt{2}} \left(\cos\psi\,
\vec{\mathbf e}_y + \sin\psi\,\vec{\mathbf e}_z\right)\nonumber \\ && \times \cos(kx-\omega t +
\phi +\psi) \nonumber \\ && - \frac{|E_+| - |E_-|}{\sqrt{2}}
\left(-\sin\psi\,
\vec{\mathbf e}_y + \cos\psi\, \vec{\mathbf e}_z\right)\nonumber \\ && \times \sin(kx-\omega t +
\phi +\psi),\label{polarized1}
\end{eqnarray}
which clearly shows that this field traces out an ellipse, whose major and minor semi-axes
have lengths $(|E_+| + |E_-|)/\sqrt{2}$ and $(|E_+| - |E_-|)/\sqrt{2}$, respectively, and
whose major semi-axis subtends the angle $\psi$ with the
$y$ axis.  The angle
$\psi$ is called the {\em polarization azimuth}, and takes on values $0\leq \psi\leq \pi$.   The
ratio between the semi-axes of the ellipse is related to
the {\em ellipticity angle} $\eta$ through the formula
$\tan\eta=\left(|E_+| - |E_-|\right)/\left(|E_+| + |E_-|\right)$.  This angle takes on values $-\pi/ 4\leq \eta\leq \pi/ 4$; see Figure~\ref{fig:ellipse}.   The sign of $\eta$ represents the direction in which the electric field $\vec{\mathcal E}(x,t)$ rotates along the perimeter of the ellipse.

Special cases of Eq.~\rf{polarized1} include linear polarization when
$\eta=0$ ($|E_+| = |E_-|$) and circular polarization when $ |\eta| =
\pi/4$.   In
particular, the field $\vec{\mathcal E}(x,t)$ is left-circularly polarized
if $\eta=-\pi/4$, that is $|E_+|=0$, and right-circularly polarized
if $\eta=\pi/4$, that is $|E_-|=0$.

To compute the angles $\eta$ and $\psi$, we proceed as follows.   First, from
Figure~\ref{fig:ellipse}, we see that
\begin{eqnarray*}&&\cos\eta=  \frac{|E_+| + |E_-|}{\sqrt {2 \left(|E_+|^2 + |E_-|^2\right)}},\\ &&\sin\eta
=
\frac{|E_+| - |E_-|}{\sqrt {2 \left(|E_+|^2 + |E_-|^2\right)}},\end{eqnarray*}
so that
\begin{subequations}
\begin{equation}
\sin2\eta=\frac{|E_+|^2 - |E_-|^2}{|E_+|^2 + |E_-|^2}.
\end{equation}
Moreover, from \rf{polarized}, we find
$E_+^*E_-=|E_+||E_-|e^{ 2i\psi}$,
which implies
\begin{equation}
\tan 2\psi = i\frac{E_+E_-^*-E_-E_+^*}{E_+E_-^*+E_-E_+^*}.
\end{equation}\label{ellipse_angles}
\end{subequations}
Formulae~\rf{ellipse_angles} provide a complete characterization of the polarization ellipse.

The concepts explained above hold equally well when the constant complex amplitude $\vec E$ is
replaced by a complex amplitude $\vec E (x,t)$ that varies slowly compared
to the plane carrier wave $e^{i(kx-\omega t)}$.  This is typically the case  for the  interaction of monochromatic light with a
$\Lambda$ configuration active optical medium~\cite{konopnicki81,Basharov90}, during which light pulses can be represented in the
form
\begin{equation} \vec{\mathcal E}(x,t)=\real\left\{\left[E_+(x,t)\,\vec {\mathbf e}_+ +
E_-(x,t)\,
\vec {\mathbf e}_- \right]e^{i(kx-\omega t)}\right\},
\label{ellipticpol}\end{equation}
where $\vec{\mathbf e}_\pm$ are the circular polarization basis vectors~(\ref{circbasis}) and $E_\pm(x,t)$ the complex envelopes of the two
circular polarization components of the light pulse inside the medium that
vary slowly compared to the wavelength and time-period of the light. As explained in the next section, the two electric field polarization components interact with the two active atomic transitions in the $\Lambda$-configuration degenerate two-level medium.  

\subsection{$\Lambda$-Configuration Maxwell-Bloch Equations}
\label{sec:mbes}

Propagation of ultra-short, monochromatic, elliptically polarized light pulses interacting resonantly with a
$\Lambda$-configuration, two-level, active optical medium, is, in the slowly-varying envelope approximation,
described by the quasi-classical system of Maxwell-Bloch equations~\cite{konopnicki81,maimistov84,Basharov90}
\begin{subequations} \label{lambdaeqns}
\begin{eqnarray}
&&\frac{\partial \hat H(t,x)}{\partial t} +\frac{\partial \hat H(t,x)}{\partial x}\nonumber  \\&& \quad =  \frac{1}{4}
\int_{-\infty}^{\infty} \;[J,\hat{\rho}(t,x,\nu)]g(\nu) d \nu,
\label{matlam1}\\
&&\frac{\partial \hat{\rho}(t,x,\lambda)}{\partial t} =
i\left[-\lambda J +\hat{H}(t,x),\hat{\rho}(t,x,\lambda) \right
].\label{Liouville}
\end{eqnarray}
\end{subequations}
Here $[\cdot ,\cdot]$ denotes the matrix commutator.

Equation (\ref{matlam1}) arises from the classical unidirectional
Maxwell's equations for the electric-field envelopes with the displacement currents on the right-hand
side.  Equation (\ref{Liouville}) is the Liouville equation for the density matrix, describing in the present case
a two-level quantum system with a doubly-degenerate ground level. The density matrix $\hat{\rho}$, the reduced Hamiltonian $\hat H$ (without the diagonal part) describing the
dipole interaction of the degenerate two-level system with the electric field, and the
matrix $J$ in Eqs. (\ref{lambdaeqns}) are
defined as
\begin{eqnarray}\label{defhrhoj}
&& \hat{\rho}=\left(\begin{array}{cccc}\displaystyle
 \mathcal N & \rho_+ & \rho_-\\
\displaystyle  \rho_+^*& n_+
& \mu  \\
\displaystyle \rho_-^*
& \mu^* & n_-\\
\end{array}\right),\quad
\hat{H}=\frac{i}{2}\left(\begin{array}{cccc}\displaystyle
 0 & E_+ & E_-\\
\displaystyle  -E_+^*& 0
& 0  \\
\displaystyle -E_-^*
& 0 & 0\\
\end{array}\right), \nonumber\\ &&
J=\left(\begin{array}{cccc}\displaystyle
 1 & 0 & 0\\
\displaystyle  0& -1
& 0  \\
\displaystyle 0
& 0 & -1\\
\end{array}\right),
\end{eqnarray}
respectively. Here, $E_\pm(x,t)$ are the
complex-valued
envelopes of the  left- and right-circular polarization components of the light pulse as given in Eq.~\rf{ellipticpol}, $\rho_\pm(x,t,\lambda)$ and $\mu(x,t,
\lambda)$ the
complex-valued
medium-polarization envelopes, and $n_{\pm}(x,t,
\lambda)$ and ${\mathcal N}(x,t, \lambda)$ the
real-valued
population densities of the degenerate ground
sub-levels and the excited level.   The electric-field
and medium-polarization envelopes, $E_\pm(x,t)$ and
$\rho_\pm(x,t,\lambda)$, are associated with the atomic transitions between
each of the ground sub-levels and the excited level, while $\mu(x,t,
\lambda)$ is the contribution to the medium polarization by the
two-photon transition between the two ground sub-levels. The
parameter $\lambda$ describes the detuning of the atomic transition
frequency from the exact resonance with the electric field, and
$g(\lambda)$ is a nonnegative function with $\int_{-\infty}^\infty g(\lambda) \,\difd\lambda =
1$ which describes the shape of the spectral line due to the inhomogeneous broadening of the atomic transitions.  The speed of
light in Eq. (\ref{matlam1}) is non-dimensionalized to $ c=1 $.
In components, Eqs. (\ref{lambdaeqns}) read
\begin{subequations}
\begin{eqnarray}
&&\frac{\partial E_{\pm}}{\partial t}+\frac{\partial E_{\pm}}{\partial x} =\int_{-\infty}^{\infty}\rho_\pm
\, g(\nu) \,d \nu , \label{erhoeqn}\\&&\frac{\partial \rho_+}{\partial t}-2i \lambda \rho_+ =  \frac{1}{2}
\left[E_+({\mathcal N}-n_+) -
E_-\mu^*\right],
\\&&\frac{\partial \rho_-}{\partial t} -2 i \lambda \rho_-=  \frac{1}{2}
\left[E_-({\mathcal N}-n_-)
- E_+ \mu \right],
\\&&
\frac{\partial \mu}{\partial t} = \frac{1}{2} \left[{E_+}^* \rho_- + E_-
{\rho_+}^* \right],
\\&&
\frac{\partial {\mathcal N}}{\partial t} = - \frac{1}{2} \left[E_+{\rho_+}^* +
{E_+}^*\rho_+ \right.\nonumber \\&&\left. + E_- {\rho_-}^* + {E_-}^*
\rho_- \right],
\\&&
\frac{\partial n_{\pm}}{\partial t} = \frac{1}{2} \left[E_{\pm}
{\rho_\pm}^* + {E_{\pm}}^* \rho_\pm\right].
\end{eqnarray}
\label{mbes}
\end{subequations}

One condition for equations (\ref{lambdaeqns}) (or (\ref{mbes})) to be valid is that the pulse-width be much shorter than the time scale of the relaxation processes in the atomic system; as discussed below, in gases, the ratio between these two time scales typically ranges from  $10^{-5}$ to $10^{-3}$~\cite{allen87}.
Also, as we already mentioned in the previous paragraph, Eq.~(\ref{matlam1})  describes unidirectional propagation.   Potential violation of unidirectionality is an important concern even for a non-degenerate two-level system, in which a spatially non-uniform density of active atoms can cause backscattering of light.   However, the most important features of  resonant interaction between light and two-level atomic systems are well-described within the unidirectional approximation~\cite{risken:4662,lamb71,lamb74,ablowitz74,kaup77,gabitov85,allen87,Basharov90}.  Since linear waves can be treated independently, bidirectionality must be taken into account only when the counter-propagating waves interact nonlinearly. Nonlinear interaction, in turn, only becomes prominent when the wave amplitudes are sufficiently large and the characteristic time of the counter-propagating waves' overlap is longer than the characteristic onset time of the nonlinear interaction.   

The amplitudes of the back-scattered waves are usually small for two reasons: the low density of the active atoms, and  the disorder in the  lower sub-level populations of the $\Lambda$-configuration system leading to randomness in the phase of the back-scattered light. 
In particular, for the typical expected value of the electric dipole,  corresponding to the resonant atomic transition, of $\sim 1$ Debye, and the resonant transition frequency $\sim 10^{15}$ sec$^{-1}$,  the density  of the active atoms $\lesssim 10^{18}$ cm$^{-3}$ induces less than 2\% of back-scattering according to the linear estimates carried out in~\cite{eilbeck72}.  (See also~\cite{Basharov90}.)    In addition, destructive summation of the back-scattered plane waves with random phases, which may result from the disorder in the  lower sub-level populations of the $\Lambda$-configuration system,  can also lead to an overall small amplitude of the back-scattered light.  Finally, we should note that in practical situations, the overlap time between two counter-propagating pulses  is very short, due to the large value of the speed of light, and therefore so is the time of the nonlinear interaction between them.

The density matrix, $\hat \rho$, is Hermitian and its time-evolution
can be represented by the formula $\hat\rho=U\hat\rho_0 U^\dagger$,
where $U$ belongs to the group $SU(3)$ and $\hat\rho_0$ is time-independent; this representation follows from Eq. (\ref{Liouville}).  Thus, the three
eigenvalues of the matrix $\hat\rho$ are conserved in time.
Alternatively, we can find three independent conserved quantities for
Eq.  (\ref{Liouville}) by computing the traces of the matrices $\hat\rho$, $\hat\rho^2$, and $\hat\rho^3$, three independent functions of the eigenvalues of $\hat\rho$.
 Explicitly, these conserved quantities are:
\begin{subequations}
\begin{equation}\label{constraint}
I_1(x,\lambda)={\mathcal N}+n_++n_-=1,\end{equation}
\begin{equation} I_2(x,\lambda)={\mathcal N}^2+n_+^2+n_-^2+2\left(\left\vert\rho_+
\right\vert^2+ \left\vert\rho_-\right\vert^2
+\left\vert\mu\right\vert^2\right),\end{equation}
\begin{eqnarray}
&&I_3(x,\lambda)={\mathcal N}^3+n_+^3+n_-^3+
3\Bigl[{\mathcal N}\left(\left\vert\rho_+
\right\vert^2+ \left\vert\rho_-\right\vert^2\right)
\nonumber \\ &&+n_+\left(\left\vert\rho_+ \right\vert^2
+\left\vert\mu\right\vert^2\right)+n_-\left( \left\vert\rho_-\right\vert^2
+\left\vert\mu\right\vert^2\right) \nonumber \\&&
+ \rho_+\rho_-^*\mu+
\rho_+^*\rho_-\mu^*\Bigr].
\end{eqnarray}
\end{subequations}
Note that unit normalization in Eq.~(\ref{constraint}) is chosen.

The $\Lambda$-configuration Maxwell-Bloch system~\rf{lambdaeqns}
contains two invariant  two-level sub-systems~\cite{allen87}, obtained by setting
either $E_+=\rho_+=n_+=\mu=0$ or $E_-=\rho_-=n_-=\mu=0$, which
describe pure two-level transitions between the excited level and
the $-$ or $+$ sub-levels, respectively.  The light involved in either of these transitions forever remains circularly polarized.

\subsection{Polarization Dynamics in a $\Lambda$-Configuration Medium}\label{sec:cauchy}

The solutions of the Maxwell-Bloch equations~\rf{lambdaeqns}
can be obtained and analyzed via the inverse scattering transform starting with the zero-curvature representation~\cite{maimistov84},
\begin{subequations}
\begin{equation}
\frac{\partial \Phi}{\partial t} = U \Phi =(i \lambda J - H)
\Phi, \label{lamlp1}
\end{equation}
\begin{equation}
\frac{\partial \Phi}{\partial x} = V \Phi =  \left(-i \lambda J + H+\frac{i}{4}
{\mathcal P}
\int_{-\infty}^{\infty}
\frac{\hat{\rho}(x,t,\nu)}{\nu - \lambda} g(\nu) d \nu \right)
\Phi,
\label{lamlp2}
\end{equation}
\label{lamlp}\end{subequations}
where the symbol $\mathcal{P}$ stands for the
Cauchy principal value of the integral, and the matrices
$H$, $\hat \rho$, and $J$ are defined in formula \rf{defhrhoj}.
 The $3\times 3$ matrix $\Phi$ is a simultaneous
solution of both Eqs.~\rf{lamlp1} and \rf{lamlp2}.  The compatibility condition of this system, $U_x-V_t+[U,V]=0$,  is equivalent to Eqs.~\rf{lambdaeqns}.

The inverse-scattering transform is well suited to address the Cauchy problem for Eqs.~\rf{lambdaeqns} formulated along the entire real axis.  Following~\cite{gabitov85}, we thus introduce the asymptotically-mixed problem in which the Cauchy data represent
 a pulse incident at the point $x=0$ and defined along the entire $t$-axis,
\begin{eqnarray}\label{incident}
&&E_\pm(t,0)=E_\pm^0(t), \qquad  -\infty<t<\infty,\nonumber\\ && \int_{-\infty}^\infty |E_\pm^0(t)|\,dt <\infty .\end{eqnarray}
The asymptotic initial state of the optical medium is given at $t=-\infty$ by \begin{subequations}\label{iconds}
\begin{eqnarray}\label{initial}
&&\lim_{t \rightarrow -\infty}\rho_\pm(x,t,\lambda)=0,\qquad \lim_{t \rightarrow
-\infty}\mu(x,t,\lambda)=0,\nonumber\\ &&  \lim_{t
\rightarrow -\infty}{\mathcal N}(x,t,\lambda)=0,
\end{eqnarray}
and
\begin{equation}
\label{alphadef}
\lim_{t \rightarrow
-\infty} n_\pm (x,t,\lambda)=\frac{1}{2}\Bigl[ 1\pm\alpha(x,\lambda)\Bigr]\geq 0,
\end{equation}
\end{subequations}
with $0<x<L$, where $L$ is the non-dimensionalized length of the sample. Here, we have assumed that only the two degenerate lower levels are populated initially.  The form of the asymptotic condition~\rf{alphadef} follows from the normalization \rf{constraint}, which also implies that $-1\leq \alpha(x,\lambda)\leq 1$.

In gases, the lifetime of the optical pulse ranges from $10^{-5}$ to $10^{-3}$ seconds, while the typical pulse-width is $10^{-8}$ seconds or shorter~\cite{allen87}.
Therefore, the above idealization of the initial time being at $-\infty$ is well justified.

The initial conditions \rf{incident} define the scattering problem at the point $x=0$ for Eq. \rf{lamlp1}, which falls in the class of Manakov's scattering problems~\cite{manakov74}.  The evolution of the scattering data in $x$ can be obtained via Eq. \rf{lamlp2},  and the electric field envelopes $E_\pm(x,t)$ can then be recovered using a set of two Marchenko-type equations~\cite{Basharov90}.  The evolution equations for the scattering data corresponding to the most general asymptotic initial state are listed in~\cite{Byrne03}; their derivation proceeds along the lines of the treatment given in~\cite{gabitov85} for the single-polarization, two-level Maxwell-Bloch equations.

As mentioned in the introduction, in~\cite{Byrne03}, using the inverse-scattering transform, a polarization switching mechanism was identified  in
the interaction of monochromatic light with a $\Lambda$-configuration optical medium initially satisfying the conditions~\rf{iconds}.  In
particular, as the pulse passes through an $x$-interval along which
$\alpha(x,\lambda)>0$ is bounded below by a positive constant, the amplitude $E_-(x,t)$ will grow in modulus towards a saturation value and $E_+(x,t)$ will decay, and vice versa if $\alpha(x,\lambda)<0$.  If the initial state of the medium is prepared using unpolarized, incoherent light, the initial population difference $\alpha (x,\lambda)$  between the two lower sub-levels can be considered a random function of $x$.  Since therefore  $\alpha (x,\lambda)$ changes sign in a random fashion, an optical pulse propagating in such a medium will experience random switching of light polarization.  Studying the statistical properties of this random switching is  the focus of this paper.

  From the inverse-scattering transform theory~\cite{ablowitz81,novikov84}, it is well known that the asymptotic behavior of the solution will be determined by the discrete spectrum of the operator in Eq.~(\ref{lamlp1})---the $N$-soliton solutions~\cite{Basharov90,Byrne03}.  In fact, in the integrable Maxwell-Bloch-type equations, if the spectral line is not infinitely narrow (i.e., $g(\lambda)\neq \delta(\lambda)$, the Dirac delta function) the continuous radiation  not only disperses away, but also becomes absorbed in the medium via Landau damping~\cite{ablowitz74}.  If the discrete spectrum of the incident pulse contains  a single eigenvalue in the upper-half $\lambda$-plane, i.e., $\lambda_1=\gamma+i\beta$, with $\beta>0$,
this pulse asymptotically reshapes itself into a single soliton ($N=1$).

We address the case when the spectral width of the pump pulse is much broader than the width of the spectral line due to the inhomogeneous broadening, $g(\lambda)$.  In this case, the initial populations can be considered homogeneous within
 the width of the spectral line, and therefore we can take
\begin{equation}\label{alphax}
\alpha(x,\lambda)=\alpha(x).
\end{equation}
The single-soliton solution is then given by the expression\begin{subequations}\label{soliton_params}
\begin{eqnarray}\label{onesoliton}
&&E_\pm(x,t)=4i\beta G_\pm(x) e^{i\Theta_\pm(x,t)}\nonumber\\
&&  \times\sech\left[2 \beta (t-x) + \tau  x
+ \frac{1}{2} \ln \frac{|d_+||d_-|}{2 \beta^2}\right.\nonumber\\ &&
+ \left. \frac{1}{2} \ln \cosh \left(2 \tau  A(x) + \ln \frac{\left\vert
d+\right\vert}{\left\vert d_-\right\vert}\right)\right],
\end{eqnarray}
where the functions
\begin{equation}\label{amp-phas}
G_\pm(x)=\sqrt{\frac{1}{2}\left[1\pm\tanh\left(2\tau A(x)+\ln\frac{\left\vert
d_+\right\vert}{\left\vert d_-\right\vert}\right)\right]},
\end{equation}
control the amplitudes of the soliton components and
the functions
\begin{equation}\label{all_phases}
\Theta_\pm (x,t) = 2 \gamma (t-x) + \sigma [x\pm A(x)]
-\arg d_\pm\end{equation}
\end{subequations}
describe their phases; the real-valued coefficients $\sigma$ and $\tau$ are given by
\begin{equation}
\label{Gdef}
 \sigma+i\tau=\frac{1}{8}
\int_{-\infty}^{\infty} \frac{g(\nu)}{\lambda_1-\nu}d \nu
\end{equation}
for the given complex number $\lambda_1=\gamma+i\beta$ with a positive imaginary part, and
\begin{equation}
\label{Adef}
A(x)=\int_0^x\alpha(\xi)\,\difd\xi,
\end{equation}
describes the cumulative initial population difference $\alpha(x)$ along the medium sample up to any given position
$x$.

Equations \rf{onesoliton} and \rf{amp-phas} show that both the electric-field components,
$E_\pm(x,t)$, of the one-soliton solution consist of the same sech-profile wave, with two
different, $x$-dependent amplitudes, proportional to the functions   $G_\pm(x)$ in Eq.~\rf{amp-phas}, respectively.  The maximal amplitude of each component equals $4\beta$.  The temporal width of the soliton
equals $1/2\beta$.  The constants $d_+$ and $d_-$ determine the phase and position of the soliton.
 If the cumulative initial population difference diverges with increasing distance into the medium, $A(x)\to\pm\infty$ as $x\to\infty$, one electric-field amplitude saturates while
the other decays, which is the one-soliton case of the polarization switching~\cite{Maimistov85,Byrne03}.

From Eq.~\rf{Gdef}, since $g(\nu)>0$ and $\lambda_1=\gamma+i\beta$ with $\beta>0$, we find for the coefficient $\tau$ the inequality
\begin{equation}\label{eq:tau}
\tau=-\frac{\beta}{8}\int_{-\infty}^\infty \frac{g(\nu)}{|\lambda_1-\nu|^2}\difd \nu  < 0 .
\end{equation}
For the Lorentzian shape of the spectral line,
\begin{equation}\label{eq:lorentzian}
g(\nu) = \frac{\varepsilon}{\pi (\varepsilon^2 + \nu^2)}
\end{equation}
the coefficients in Eq.~\rf{Gdef} become
\begin{equation}
\label{Glorentz}
 \sigma=  \frac{\gamma}{8\left[\gamma^2 + (\beta + \varepsilon)^2\right]}, \quad \tau=-\frac{\beta + \varepsilon}{8\left[\gamma^2 + (\beta + \varepsilon)^2\right]}. \end{equation}
   Note that $\sigma=0$ in Eq.~\rf{Gdef} if $\gamma=0$, which is also easily shown to be true for any even
spectral line shape $g(\nu)$. 

The soliton speed and the phases of its components depend on the position $x$ along the optical medium.  Here, we compute the soliton speed as follows:
At each position $x$, both components of the soliton reach their peak  intensity at the time
for which the argument of the sech-profile vanishes.  This condition and Eq.  \rf{onesoliton}
 give
the travel time of the soliton from when it is injected into the medium at $x=0$ until it reaches the point $x$
as
\begin{eqnarray}\label{Teq} &&\st(x)=x-\frac{1}{2\beta}\left[\tau x +\frac{1}{2} \ln\frac{2\left\vert
d_+ d_-\right\vert}{\left\vert
d_+ \right\vert^2+\left\vert
 d_-\right\vert^2} \right. \nonumber\\ &&\left. + \frac{1}{2} \ln \cosh
\left(2 \tau  A (x) + \ln \frac{\left\vert
d+\right\vert}{\left\vert d_-\right\vert}\right) \right].
\end{eqnarray}
From Eq.~(\ref{Teq}), it is clear that the
speed of the soliton thus satisfies the equation
\begin{eqnarray}\label{velocity}
&&\frac{1}{v_{\rm soliton}(x)}=\frac{\difd\st}{\difd x}(x) = 1- \frac{\tau }{2\beta} \biggl[ 1+
\alpha (x) \nonumber\\ && \times \tanh \left(2 \tau  A(x) + \ln \frac{\left\vert
d+\right\vert}{\left\vert d_-\right\vert}\right)\biggr] .
\end{eqnarray}
Using Eq.~\rf{eq:tau}
and the fact that $ |\alpha (x) | \leq 1 $,
we readily conclude that $v_{\rm soliton}(x)\leq 1$, i.e, that the soliton speed never exceeds the speed of light.
If the initial population difference $\alpha$ is $x$-independent, the soliton velocity asymptotically behaves as
\[v_{\rm soliton}(x\to\infty)\to \left[1+ \frac{|\tau |(1-|\alpha|)}{2\beta}\right]^{-1}.\]

To compute the polarization azimuth $\psi$ and the angle of ellipticity $\eta$
for the one-soliton solution, we insert the components of the solution~\rf{onesoliton} into Eqs.~\rf{ellipse_angles}
to obtain
\begin{subequations}\label{psieta}
\begin{eqnarray}\label{psi1}
&&\psi = -\sigma A(x) +\frac{1}{2}\arg \left(d_-^* d_+\right), \\&& \sin 2\eta =
\tanh \left( 2\tau A(x) +
\ln\frac{\left\vert d_+\right\vert}{\left\vert
d_-\right\vert}\right).\label{eta1}
\end{eqnarray}
\end{subequations}
Note that these two angles are independent of the time $t$: at any point $x$ along the medium,
the light polarization remains constant in time as the soliton passes by.

Note also that the polarization azimuth $\psi$ remains constant if the parameter $\sigma$ vanishes.  Recalling that the eigenvalue $\lambda_1$ is the complex number $\lambda_1=\gamma+i\beta$ and the remark after Eq.~\rf{Glorentz}, we see that this happens when $\gamma=0$ and so $\lambda_1=i\beta$, i.e., pure imaginary, provided that the spectral line shape $g(\lambda)$ is an even function.  From Eqs.~\rf{soliton_params}, its is easy to see that this case contains solitons with real-valued electric field components, which are obtained with the appropriate choice of the constants $d_\pm$.  In other words, when the spectral line shape $g(\lambda)$ is an even function, the polarization azimuth $\psi$ of all solitons with real-valued electric field components remains constant, and so is independent of the distance $x$ into the medium.

\section{Soliton dynamics in the presence of  spatial disorder in the medium population}
\label{population:disorder}

We now describe light propagation in the presence of spatial disorder in the initial  population densities, characterized by the function   $\alpha(x)$ in Eqs.~\rf{alphadef} and~\rf{alphax}. The spatial distribution of the initial population is determined by the manner in which the atomic system is prepared. In general, for the $\Lambda$-configuration with two degenerate levels,  it is difficult to control the relative populations of the sub-levels during the preparation process. For example, if the system is prepared using unpolarized or partially polarized pump light, the relative distribution of the sub-level populations will be random.  We will be concerned with how this randomness induces random polarization switching in the one-soliton solution~\rf{onesoliton}.

\subsection{White Noise Approximation to the Initial Population Density Difference}
\label{sec:wienercoarse}
We assume the initial population density difference $\alpha(x)$ in the medium to be random and spatially homogeneous in the statistical sense, and treat it as homogenous white noise with amplitude $a$ superposed upon a mean (bias) $b$:
\begin{subequations}
\begin{eqnarray}\label{eq:mean}
&&\langle \alpha (x) \rangle  = b,\\ \label{eq:corfun}
&&\langle [\alpha (x) - b] [\alpha (\xp)-b]
 \rangle = a^2\delta(x-\xp),
 \end{eqnarray}
 \label{eq:whitemodel}
\end{subequations}
where $\langle\cdot\rangle$ denotes ensemble averaging over the statistical ensemble of all possible realizations of the initial population difference $\alpha(x)$, and $\delta(\cdot)$ is the Dirac delta function.

The white-noise characterization \rf{eq:whitemodel} of the initial population density difference $\alpha(x)$ is consistent with the Maxwell-Bloch model~\rf{lambdaeqns}  when the correlation length $\lcor$ of $\alpha(x)$ (discussed in more mathematical detail in Appendix~\ref{sec:lcor}) satisfies three conditions.    The first is that
\begin{equation}\label{eq:lcormorelambda}
\lcor\gg\lambda_0,
\end{equation}
where $\lambda_0$ is the wavelength of the light interacting resonantly with the transitions between the ground and excited levels in the $\Lambda$-configuration medium under investigation. The second is
is that  $L_c$ should be much shorter than the typical spatial pulse-width,
\begin{equation}\label{eq:lcorlessbeta}
\lcor\ll\frac{1}{\beta}.
\end{equation}
The third condition is
\begin{equation}\label{eq:lcorlessx}
\lcor\ll x,
\end{equation}
where $ x $ is the position of the observation point along the medium.

As explained below,  conditions \rf{eq:lcormorelambda} and \rf{eq:lcorlessbeta} are necessary to make the modeling of the initial population difference $\alpha(x)$ by random noise compatible with the Maxwell-Bloch equations~\rf{lambdaeqns} (or \rf{mbes}).  Condition~\rf{eq:lcorlessx} is what allows the approximation of the true initial population difference $ \alpha (x) $ by the idealized white-noise model~\rf{eq:whitemodel}.   In fact, condition~\rf{eq:lcorlessx} follows from condition~\rf{eq:lcorlessbeta} in any realistic experimental device, which would be long compared to the soliton width $ 1/\beta $.
We now proceed to discuss the need and consequence of these three conditions in more detail.

Condition~\rf{eq:lcormorelambda} must hold because Eqs.~\rf{lambdaeqns} (or \rf{mbes}) describe slowly-varying envelopes of the electric field and medium polarization components, and their carrier-wave oscillations are averaged out in the process of deriving these equations.   The correlation length $\lcor$ must therefore be sufficiently large compared to the wavelength $\lambda_0$ of the light interacting with the medium not to be averaged out as well.  In other words, in order for the envelope approximation leading to Eqs.~\rf{lambdaeqns} to be valid simultaneously with  the assumption~\rf{eq:corfun}, we must assume condition~\rf{eq:lcormorelambda}.

Condition~\rf{eq:lcorlessbeta} should hold because Eqs.~\rf{lambdaeqns} (or \rf{mbes}) employ the approximation of unidirectionality.   If the correlation length $L_c$ of the medium with the initial population difference $\alpha(x)$ was comparable to or larger than the spatial width of the light pulse traveling through this medium, this random population difference could induce considerable backscattering of the pulse.   Consequently, the pulse could be destroyed and the unidirectionality approximation violated.  (Cf. the detailed discussion of a similar problem in~\cite{Chertkov01,ChertkovGabitovLushnikovMoeserToroczkai2002}.)

To understand the meaning of the condition~\rf{eq:lcorlessx}, let us recall that Eqs.~\rf{soliton_params} describing the soliton only involve the population difference $\alpha(x)$ through its cumulative spatial effect expressed by its spatial integral, $A(x)$ in Eq.~\rf{Adef}.  In particular, when~\rf{eq:lcorlessx} holds, the integral $A(x)$ can be well-approximated as \begin{equation}
\label{Amodel}
A(x)= aW(x)+bx,
\end{equation}
where $W(x)$ is the usual Wiener process~\cite{breiman92}.   Recall that the Wiener process $W(x)$ is for each $ x $ a mean-zero Gaussian  random variable with variance $ x $ and probability density function
\begin{equation}\label{gaussian}
p_W(s;x)=\frac{1}{\sqrt{2\pi x}}\exp\left(-\frac{s^2}{2x}\right),
\end{equation}
where $s$ parametrizes the range of the random variable $W(x)$.   The representation \rf{Amodel}, which is equivalent to \rf{eq:whitemodel}, also applies to Eqs. (\ref{Teq}) and (\ref{psieta}) for the soliton travel time and ellipticity angle, respectively.

The parameter $ a $ in the approximation~\rf{eq:whitemodel} (or~\rf{Amodel}) is related to 
 the correlation length $\lcor$,and variance $\sigma^2_\alpha$ of any given physical  initial population difference $\alpha(x)$ through the equation
$ a = \sqrt{2\lcor}\, \sigma_\alpha$, as  follows from the discussion in Appendix~\ref{sec:lcor}.

We should remark, however, that the white noise approximation~\rf{eq:whitemodel} does not make literal sense, as it violates the constraint $ |\alpha (x) |\leq 1 $ implied by the normalization~(\ref{constraint}). More generally, it is not a valid description of physical quantities that depend on the initial population difference $\alpha(x)$ itself, such as the local soliton speed in Eq.~\rf{velocity}.   The precise description of such quantities would require a more detailed model of $\alpha(x)$, which we do not pursue here.
However, state variables such as polarization variables and travel time, which involve the integrated  effects of $ \alpha (x) $, have their statistics well described by the white noise approximation~\rf{eq:whitemodel} under the asymptotic conditions 
(\ref{eq:lcormorelambda}) through (\ref{eq:lcorlessx}).

In an experiment, the correlation length $L_c$ of the population difference $\alpha(x)$ would be approximately the same as the coherence length $\ell_c$ of the pump light used to prepare the optical medium.   The related characteristic dephasing time $t_d=\ell_c/c$, where $c$ is the speed of light (set to unity in our dimensionless coordinates),  is determined by the width $\Delta \nu_p$ of the spectral line of the light source as $t_d \sim 1/\Delta \nu_p$. Therefore, the coherence length $\ell_c$ is determined as
\[
\ell_c \sim c t_d = \frac{c}{\Delta \nu_p}.
\]
Taking into account that the wavelength $\lambda_p$ is related to the frequency $\nu_p$ as   $\lambda_p =c/ \nu_p$, we obtain \[\frac{d \nu_p}{d \lambda_p}=-\frac{c}{\lambda_p^2}\] and so, considering just the magnitudes of $\Delta\nu_p$ and $\Delta\lambda_p$, we find that \[\Delta \nu_p =\frac{c}{\lambda^2_p}\Delta \lambda_p.  \] From this formula, it finally follows that the coherence length $\ell_c$ is determined as
$$
\ell_c \sim \frac{\lambda_p^2}{\Delta \lambda_p}.
$$
Here, $\lambda_p$ is the average wavelength of the pump light and $\Delta \lambda_p$ is the characteristic width of the pump light-source spectral line (in terms of wavelength).

To demonstrate experimental feasibility, we recall that
for Ti-sapphire lasers, for example, these parameters are $\Delta \lambda_p \sim 5 nm$ and $\lambda_p \sim 800nm$, therefore $\ell_c \sim 10^5 nm = 0.1 mm $~\cite{wolf07}.  For a typical $\Lambda$-configuration transition pair in the visible regime (e.g., in sodium vapor, the wavelength corresponding to the transition is $\lambda_0\sim 600nm$), this argument shows that $\lambda_0\ll \ell_c\sim \lcor$ on the one hand, and that a
several-centimeters long experimental device
is clearly sufficiently long  to capture the desired statistical effects. 
Therefore, both conditions~\rf{eq:lcormorelambda} and \rf{eq:lcorlessx} can be satisfied simultaneously in this case.

Noting that the soliton travel time $ \st (x) $ and the ellipticity angle $ \eta (x) $ depend on the initial population difference only through the 
product $ \tau  A(x) $, 
defined in Eqs.~\rf{Gdef} and \rf{Adef}, respectively, we here identify three fundamental length scales associated with the dynamics of this quantity, and thus, through $ \eta (x) $, also the polarization switching.
First, as seen from Eq.~\rf{Amodel},
\begin{equation}\label{eq:lb}
L_b=\frac{1}{|\tau | |b|}
\end{equation}
is the length scale over which the deterministic bias $b$ in Eq.~\rf{eq:mean} induces a significant change in 
$ \tau  A(x) $.
Second,
\begin{equation}\label{eq:la}
L_a = \frac{2}{a^2 \tau^2 }
\end{equation}
is the length scale over which the random component of the initial population difference fluctuations in the medium, given approximately as $aW(x)$ in Eq.~\rf{Amodel}, creates a signficant change in $ \tau  A(x) $.  Finally,
\begin{equation}\label{eq:lfluc}
\Lab = \frac{\Lb^2}{\La}= \frac{a^2}{b^2},
\end{equation}
is the length scale before which random fluctuations dominate the effects of the deterministic bias, and after which the opposite is true.  Because of Eq.~\rf{eq:lfluc}, these three length scales must obey one of the two following orderings:
\begin{align}
\La \leq \Lb \leq \Lab & \quad\text{or}\quad \Lab \leq \Lb \leq \La. \label{eq:labord}
\end{align}
Note that
\begin{subequations}
\begin{align}
&L_b\to \infty,\quad  \Lab \to \infty  \quad&\mbox{as}\quad b\to 0,\\
&L_a\to \infty,\quad \Lab \to 0 &\mbox{as}\quad a\to 0.
\end{align}
\end{subequations}
Note also that the lengths $L_a$ and $L_b$ depend on the initial medium parameters $a$ and $b$ in Eqs.~\rf{eq:whitemodel} as well as the soliton parameter $\tau$ in Eq.~\rf{Gdef}, while $\Lab$ only depends on $a$ and $b$.

\begin{figure}[h]\begin{center}
\includegraphics[scale=.4]{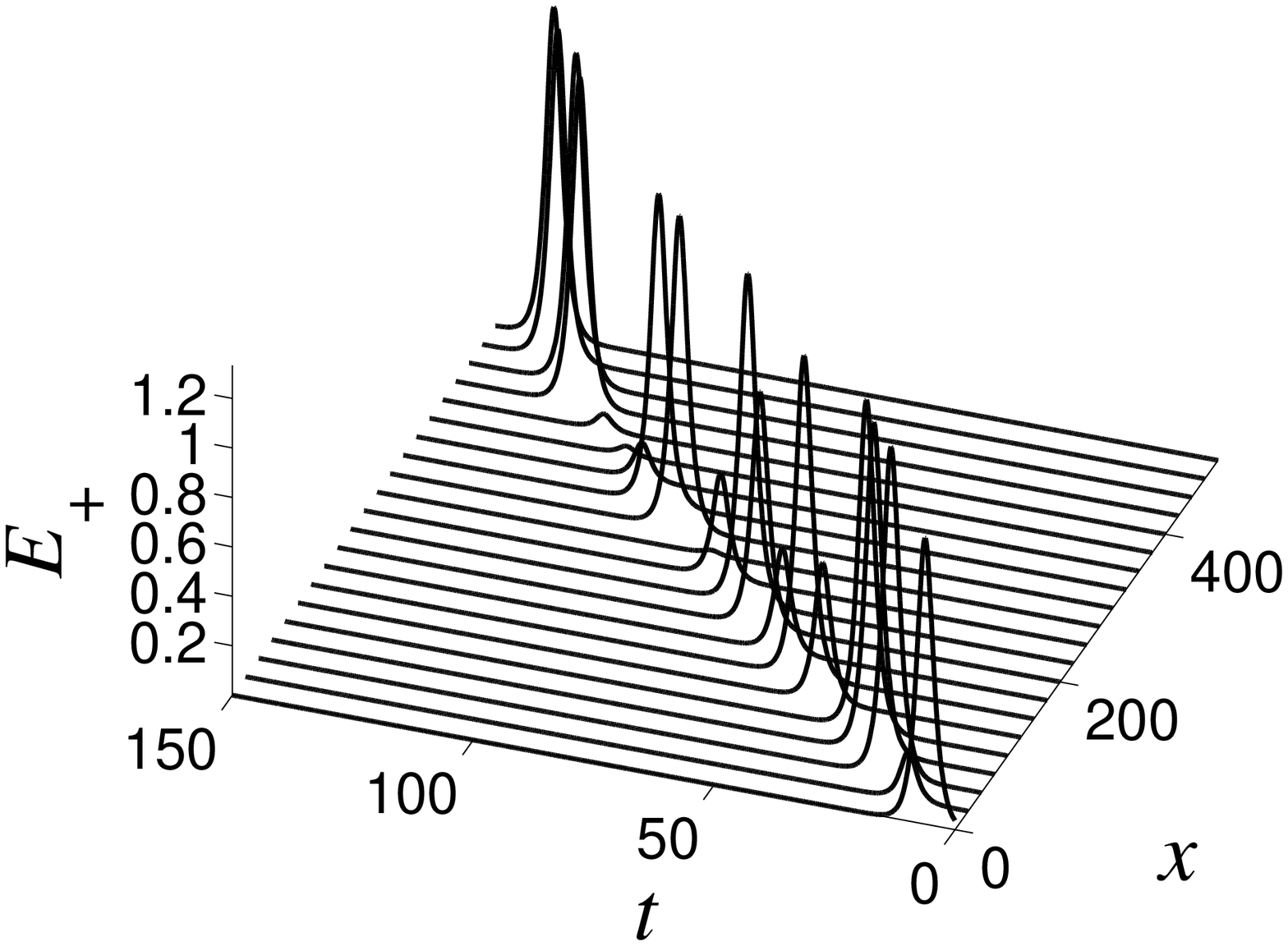}
\includegraphics[scale=.4]{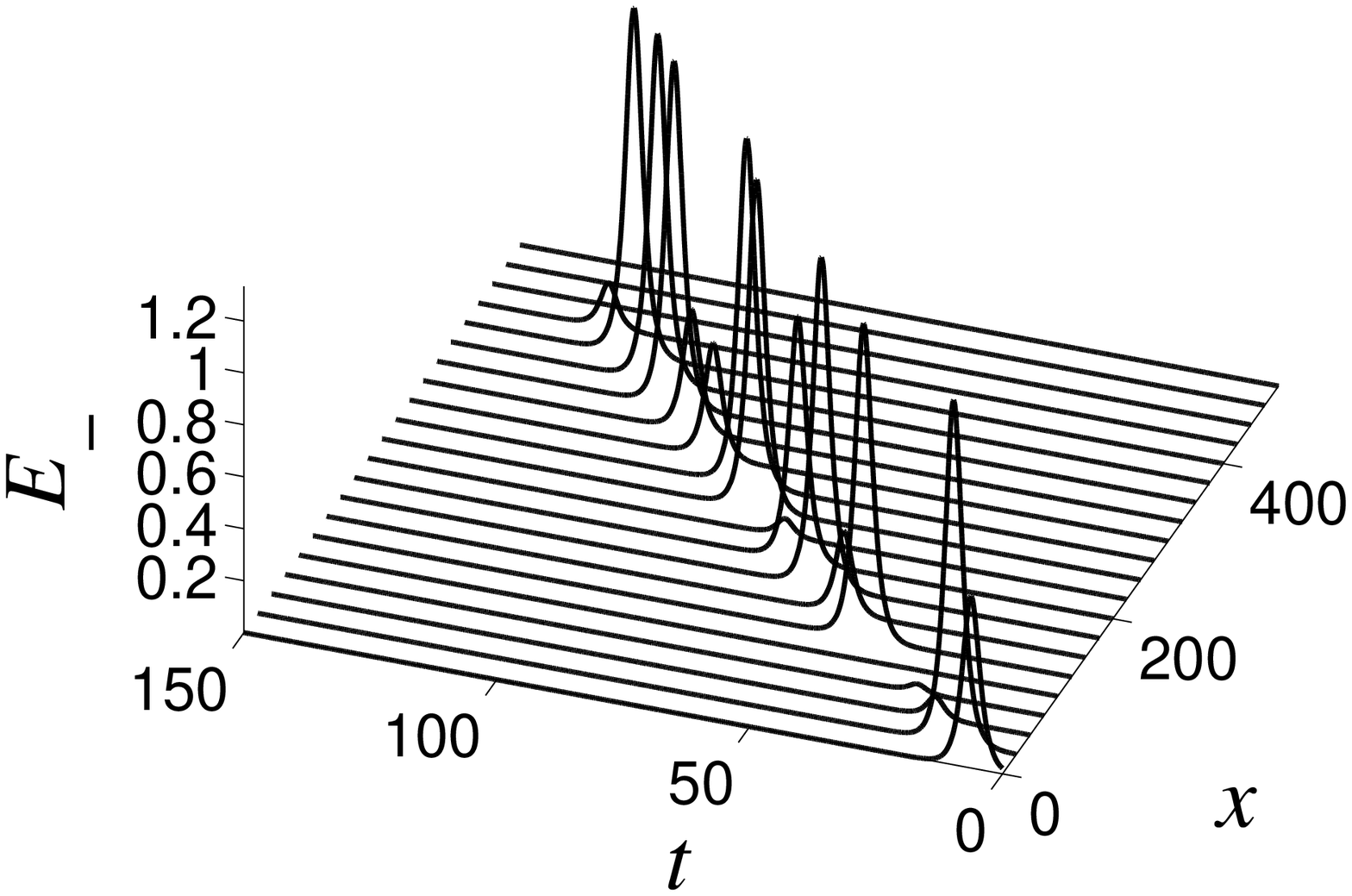}
\caption{\label{fig:solitons} The two circular-polarization amplitudes, $E_+$ and $E_-$, of a real-valued, polarized-light soliton pulse propagating through a randomly-prepared $\Lambda$-configuration optical medium, as described by Eq.~(\ref{onesoliton}).   The model of the initial population difference $\alpha(x)$ was sampled from the properly rescaled beta distribution with mean $b=\langle \alpha(x)\rangle = 0.001$,  variance $\sigma_\alpha^2=0.71$, and coherence length $\lcor=1.8$.  The width of the Lorentzian spectral line in Eq.~\rf{eq:lorentzian} is $\varepsilon=0.1$. The soliton parameters are  $\beta=1/3$, $\gamma=0$, $d_+=d_-=i$. }\end{center}
\end{figure}

The polarization azimuth $ \psi (x) $ depends on the disorder of the initial medium occupation numbers through the product $ \sigma A(x) $ which, as we will see in Sec.~\ref{par:azimuthstats}, does not require length scales analogous to $ L_a $ and $ L_b $.

The question that we need to answer is how the polarization of the pulse behaves at large distances $x$ into the medium. If the initial difference $\alpha(x)$ between the populations of the degenerate ground states of the medium exhibits a bias, $\langle \alpha(x)\rangle = b$, it is reasonable to expect that the soliton will eventually evolve into a single circular polarization, which will depend on the sign of this bias.  If no such bias exists, that is, $\langle \alpha(x)\rangle = b =0$, then it is reasonable to expect that
the soliton will switch intermittently between left and right
circular polarizations over large distances.   In the forthcoming sections, we will confirm this intuition explicitly.

\subsection{Soliton Statistics at Fixed Observation Point}
\label{sec:wienerstats}

In this section, we calculate the  statistics of the soliton travel time $\st(x)$ in Eq.~(\ref{Teq})
and the two angles that determine the dynamics of the polarization ellipse, i.e., the polarization azimuth $\psi$ and angle of ellipticity $\eta$  in Eqs.~(\ref{psi1}) and (\ref{eta1}).  Throughout the section, we employ the white noise approximation \rf{eq:whitemodel} of the initial population density difference $\alpha(x)$ (or, equivalently, the Wiener process approximation~\rf{Amodel} for its spatial integral $A(x)$). From the previous section, we recall that this requires the observation point $x$ to be sufficiently far into the medium in comparison with the correlation length $\lcor$ of the function $\alpha(x)$ (cf. Eq.~\rf{eq:lcorlessx}).

\subsubsection{Soliton Travel Time}
\label{sbsb:travel}

As we recall from Section~\ref{sec:cauchy}, the soliton travel time $\st(x)$, given by Eq. (\ref{Teq}), is the time needed for the peak of the soliton to reach the observation position $x$.
The time $\st(x)$ is a random function with the randomness arising solely from the integral $A(x)$ of the initial population density difference $\alpha(x)$, defined in Eqs.~\rf{Adef}, \rf{alphadef}, and \rf{alphax}, respectively.  As in Section~\ref{sec:wienercoarse}, we assume the Wiener process representation~\rf{Amodel} for $A(x)$, i.e., $A(x)=aW(x)+b$.   In the calculations below, for each fixed $ x $, we parameterize the range of the random variable $ W(x) $  by the variable $ s $.

\paragraph{Mean and Variance of the Soliton Travel Time}
The mean and variance of the soliton travel time $\st(x)$ can be expressed as the integrals
\begin{subequations}\label{eq:tstats}
\begin{align}
\langle \st(x)\rangle=&  \int_{-\infty}^\infty \frac{\tilde{\st}(x,s)}{\sqrt{2\pi x}}\exp\left(-\frac{s^{2}}{2x}\right)\,\difd s,
\label{expectation}\\
\sigma^2_\st (x)=&\int_{-\infty}^\infty \frac{\left[\tilde{\st}(x,s)-\langle \st(x)\rangle\right]^{2}}{\sqrt{2\pi x}}\exp\left(-\frac{s^{2}}{2x}\right)\,\difd s.
\end{align}
\end{subequations}
where $ \tilde \st (x,s) $ is defined as   $\st(x)$ in Eq.~(\ref{Teq}) with $A(x)$ replaced by $as+bx$.

In general, the integrals in Eqs.~\rf{eq:tstats} can only be evaluated numerically.  For sufficiently large distances, however, we can exploit the formula
\begin{equation}\label{logcosh}
\ln \cosh (u) =  |u|-\ln 2 +  O(\expe^{-2|u|}),
\end{equation}
valid for $|u|\gg 1$,
in Eq.~(\ref{Teq}), to approximate $\tilde \st(x,s)$ in Eqs.~\rf{eq:tstats} as
\begin{eqnarray}\label{tstapprox}
&&\tilde\st(x,s)\sim x-\frac{1}{2\beta}\left[ \tau x + \frac{1}{2} \ln\frac{|d_+d_-|}{|d_+|^2+|d_-|^2} \right.\nonumber\\  &&\left.+ \frac{1}{2}\left|2\tau(as + bx) + \ln\frac{|d_+|}{|d_-|}\right| \right]
\nonumber\\  &&+ O(\expe^{-|2 \tau (a s + bx)|}).
\end{eqnarray}
In this and the following asymptotic statements, we will treat $ \ln (|d_+|/|d_-|) $ as a fixed quantity of order unity (which, in connection with Eq.~(\ref{eta1}), means the pulse polarization at the entrance to the medium is not close to circular);  otherwise, additional length scales involving $ \ln (|d_+|/|d_-|) $ would appear in the error estimates.
The mean soliton travel time deep into the medium can then be expressed as
\begin{widetext}
\begin{subequations}\label{eq:tstatsapprox}
\begin{equation}\label{apprexpectation}\langle \st(x) \rangle \sim \begin{cases}
\displaystyle x-\frac{1}{2\beta}\left[ \tau x(1-|b|) +\frac{1}{2}  \ln\frac{|d_-|^2}{|d_+|^2+|d_-|^2}\right] &\\[12pt]
 + O \left(\expe^{-x/\Lb} + (x/\Lb), \expe^{-x/\Lab}\right), & \text{ for } x \gg \Lb, \Lab,\\[12pt]
\displaystyle x-\frac{1}{2 \beta}\left[\tau x + \sqrt{\frac{2}{\pi}} |\tau a | \sqrt{x} + \frac{1}{2} \ln \frac{|d_+d_-|}{|d_+|^2+|d_-|^2}\right]  &  \\[12pt]
+ O \left((x/\La)^{-1/2} + (x/\La)^{1/2} (x/\Lab)\right), & \text{ for } \La \ll x \ll \Lab,
\end{cases}
\end{equation}
and its variance as
\begin{equation} \label{eq:tstvarapprox}
\sigma^2_\st(x)\sim
\begin{cases}
\displaystyle \frac{\tau^2  a^{2}x}{4\beta^{2}} + O \left(\expe^{-2x/\Lb} + (x/\Lb)^2 \expe^{-x/\Lab}\right), &\text{ for } x \gg \Lb, \Lab, \\[12pt]
\displaystyle  \frac{(\pi-2)\tau^2  a^{2}x}{4\pi\beta^{2}}-\frac{4- \pi}{16 \pi \beta^{2}}
\left(\ln\frac{|d_+|}{|d_-|}\right)^{2} &\\[12pt]
 + O\left((x/\La)^{-1/2}+ (x/\La) (x/\Lab)^{1/2}\right), & \text{ for } \La \ll x \ll \Lab,
\end{cases}
\end{equation}
\end{subequations}
\end{widetext}
where the parameter $\tau$ is defined in Eq.~(\ref{Gdef}).

Note that since $\tau<0$ due to Eq.~\rf{eq:tau}, the expectation value of the soliton travel time $\st(x)$ in Eq.~\rf{apprexpectation}  increases linearly for large observation-point distance $x$.
Recalling from Eqs.~\rf{eq:lb}, \rf{eq:la}, and~\rf{eq:lfluc} that the regime $ x \ll \Lab $ is dominated by the random fluctuations in the initial population difference, whereas the drift due to the bias $b$ in the difference dominates at length scales $ x \gg \Lab $, we note that the first line in each display corresponds to the drift-dominated case while the second line corresponds to the noise-dominated case.   In particular, the case $x \gg \Lb, \Lab$ only corresponds to nonzero average initial population density difference $b=\langle\alpha(x)\rangle$, while the case $\La \ll x \ll \Lab$ also contains the case $b=0$.  Because of the length-scale relationships (\ref{eq:labord}), the regimes considered provide a comprehensive description for the soliton travel time deep in the medium.

\paragraph{Probability Distribution of the Soliton Travel Time}  The cumulative distribution function
\begin{equation}\label{cumuleq}
F_{\st}(t;x)=\Prob \Bigl\{\st(x) \leq t\Bigr\}.
\end{equation}
for soliton travel time, $\st(x)$ in Eq.~\rf{Teq}, from the entrance of the medium to a given position $ x $,  can be computed in the Wiener process approximation using Eqs.~\rf{Teq} and \rf{Amodel}, which yield the expression
\begin{align*}
F_{\st}(t;x)&= \Prob \Bigl\{W(x)\leq u_-(x,t)\text{ or } W(x) \geq u_+ (x,t)\Bigr\} \\[8pt]
&= \min\Biggl[1,\Prob\Bigl\{W(x)\leq u_-(x,t)\Bigr\}\\
& + \Prob\Bigl\{W(x)\geq u_+ (x,t) \Bigr\}\Biggr]
 \end{align*}
where
\begin{equation}\label{upm}
u_\pm(x,t)=\frac{1}{2a\tau}\left(-2b\tau x-\ln\frac{\vert d_+\vert}{\vert d_-\vert} \mp \cosh^{-1}\kappa (x,t)\right)
\end{equation}
and
\begin{equation}\label{kappaxs}
\kappa(x,t)=\exp\left(-4\beta (t-x) -2\tau x - \ln \frac{2\vert d_+d_-\vert}{|d_+|^2+|d_-|^2} \right).
\end{equation}
Since $W(x)$ is a normally distributed random variable with mean 0 and variance $x$, we find
\begin{equation}\label{timecum}
F_{\st}(t;x)=
\begin{cases}
\displaystyle 1+ \frac{1}{2}\erf\left(\frac{u_-(x,t)}{\sqrt{2x}}\right) &  \\[12pt]
\displaystyle  - \frac{1}{2}\erf\left(\frac{u_+(x,t)}{\sqrt{2x}}\right), & \text{ for }
0 < t < \tmax (x), \\[8pt]
1, & \text{ for } t \geq \tmax(x),
\end{cases}
\end{equation}
where
$\tmax (x)$ is the upper bound on the soliton travel time to a position $ x $, given by
\begin{displaymath}
\tmax (x) = \left(1 - \frac{\tau}{2\beta}\right)x -\frac{1}{4 \beta}
 \ln \frac{2\vert d_+d_-\vert}{|d_+|^2+|d_-|^2},
 \end{displaymath}
 and
the error function is defined as
\begin{equation}\label{error}
\erf (y) = \frac{2}{\sqrt{\pi}} \int_{0}^{y} \exp\left(-z^2\right) \, dz.
\end{equation}

The probability density function $p_{\st}(t;x)$  of the soliton travel time $\st(x)$ is given by the partial
derivative of the cumulative distribution function $F_{\st}(t;x)$ with respect to $t$.  After differentiation of Eq.~\rf{timecum} and some algebra, we arrive at the expression
\begin{widetext}
\begin{equation}
p_{\st}(t;x)= \\
\begin{cases}
\displaystyle \sqrt\frac{2}{\pi x}\;\frac{\beta\,\kappa(x,t)}{a|\tau| \sqrt{\kappa^2(x,t)-1}}\left[\exp\left(-\frac{u_+^2(x,t)}{2x}\right)+\exp\left(-\frac{u_-^2(x,t)}{2x}\right)\right], \quad &  \text{for } 0 \leq t \leq \tmax (x), \\[12pt]
0, & \text{otherwise},
\end{cases}
\end{equation}
\end{widetext}
with $u_\pm(x,t)$ and $\kappa(x,t)$ as in Eqs.~\rf{upm} and~\rf{kappaxs}, respectively.
\begin{figure}
\includegraphics[scale=.4]{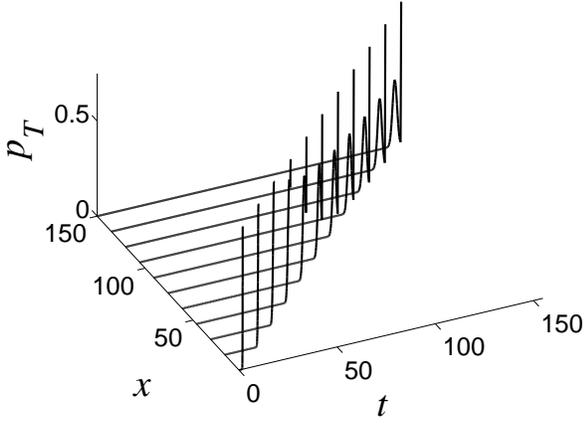}
\caption{Probability density function, $p_\st(t;x)$, with $\beta=1$, $\gamma=1$,  $\varepsilon=0.1$, $d_+=i$, $d_-=i/3$,  $a=3$, $b=0.8$.  }
\label{fig:pt}
\end{figure}

A sample plot of the probability density function $p_{\st}(t;x)$ of the soliton travel time $\st(x)$ is presented in Figure~\ref{fig:pt}.  Note the local maximum emerging from the boundary  at $t=\tmax (x)$
at large values of $x$, i.e., locations deep into the medium sample.

\subsubsection{Polarization Variables}
\label{sbsb:polarstats}

To describe the statistics of the polarization azimuth $\psi$ and the angle of ellipticity $\eta$, we again use the
Wiener process approximation~\rf{Amodel} and replace the function $ A(x) $ in the expressions (\ref{psieta}) for the polarization variables with $ a W(x) + bx $.  The statistics can then be obtained as follows.

\paragraph{Polarization Azimuth Statistics}
\label{par:azimuthstats}
As the polarization azimuth $ \psi (x) $ is a linear function of the Wiener process $ W(x) $, it itself behaves like
a Brownian motion with drift $ - \sigma  b $ and diffusion coefficient $ \frac{1}{2} \sigma ^2 a^2 $.  That is, its probability density at any position $ x $ is given by a Gaussian form
\begin{eqnarray}
&& p_{\psi}(s;x)=\frac{1}{\sqrt{2\pi x}\,a\sigma}\nonumber \\ && \times \exp\left \{-\frac{\Bigl[s-\frac{1}{2}\arg(d^*_-d_+)+\sigma bx\Bigr]^{2}}{2x a^{2}\sigma^{2}(\lambda)}\right\}.
\end{eqnarray}
with mean
\begin{equation}
\langle\psi (x)\rangle= -\sigma bx +\frac{1}{2}\arg(d^*_-d_+)
\end{equation}
and variance
\begin{equation}
\sigma^2_\psi (x) =\sigma^{2}a^{2}x.
\end{equation}

Note that when $\sigma=0$,
\[ p_{\psi}(s;x) = \delta\left( s-\frac{1}{2}\arg(d^*_-d_+) \right), \] where $\delta(\cdot)$ is the Dirac delta function, i.e., the dynamics of the polarization azimuth becomes constant, as was mentioned at the end of Section~\ref{sec:polarization_dyn}.  In particular, the polarization azimuth is constant for all solitons whose electric-field envelopes are real-valued, so that for such solitons, the polarization ellipse does not rotate.

\begin{figure}
\includegraphics[scale=.4]{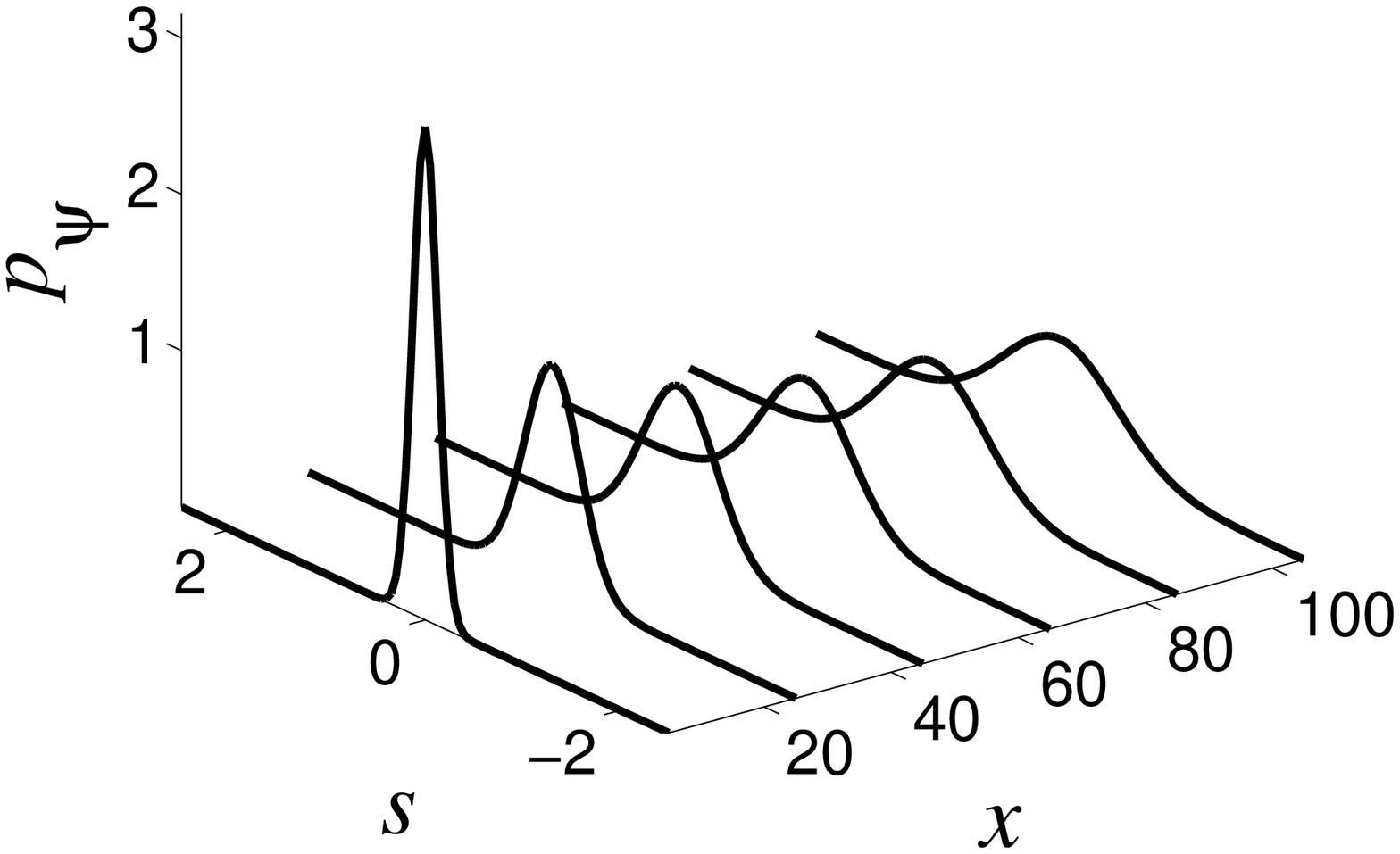}
\includegraphics[scale=.4]{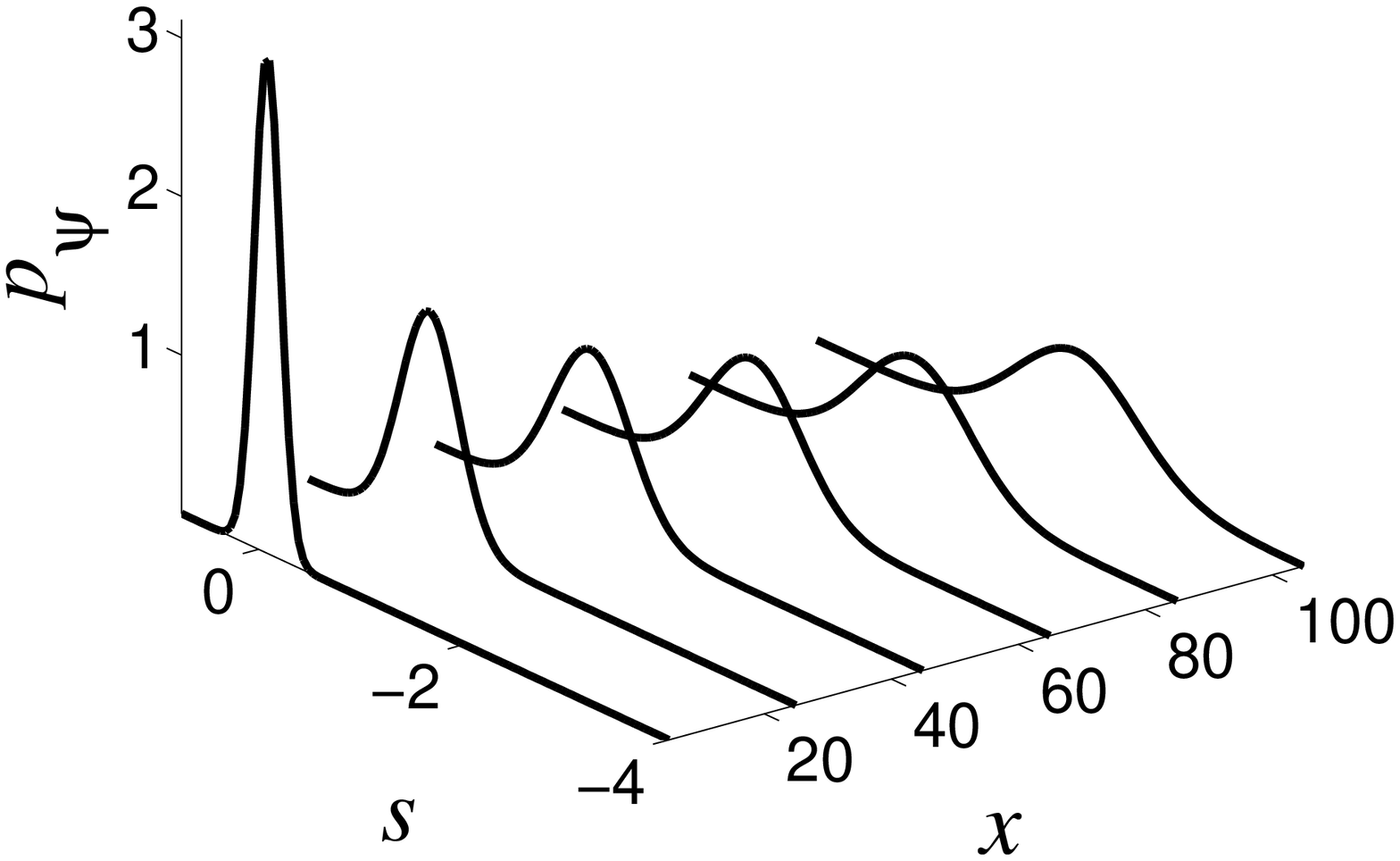}
\caption{Probability density function, $p_\psi(s;x)$, with $\beta=1$, $\gamma=1$,  $\varepsilon=0.1$, $d_+=i$, $d_-=i/3$,  $a=1$.  Top: no bias, $b=0$.  Bottom: nonzero bias, $b=0.3$. }
\label{fig:ppsi}
\end{figure}

\paragraph{Mean and Variance of the Ellipticity Angle}
\label{sec:meaneta}
For the angle of ellipticity $\eta(x)$, just as for the soliton travel time $\st(x)$, the initial medium population difference $\alpha(x)$ again contributes to the evolution of $\eta(x)$ in the $x$ direction through the 
Wiener process $\tau A(x)$.
 Therefore, the length scales given by Eqs.~\rf{eq:lb}, \rf{eq:la}, and~\rf{eq:lfluc} are again relevant here.

Taking $\Eta(x,s)$ to be defined as the expression for $\sin2\eta$ in Eq.~(\ref{eta1}) with $A(x)$ replaced by $as+bx$, i.e.,
\begin{equation}
\Eta(x,s) =
\tanh \left( 2\tau (as+bx) +
\ln\frac{\left\vert d_+\right\vert}{\left\vert
d_-\right\vert}\right),
\end{equation}
the expectation and variance of $ \sin 2\eta$ can be expressed as
\begin{subequations}\label{eq:etameanvar}
\begin{equation}
\langle \sin 2\eta  \rangle =
\langle \Eta (x,W(x)) \rangle = \int_{-\infty}^\infty \frac{\Eta(x,s)}{\sqrt{2\pi x}}\exp\left(-\frac{s^{2}}{2x}\right)\,\difd s
\end{equation}
\begin{equation}
\sigma^2_{\sin 2\eta}=\int_{-\infty}^\infty \frac{(\Eta(x,s)-\langle \sin(2\eta)\rangle)^{2}}{\sqrt{2\pi x}}\exp\left(-\frac{s^{2}}{2x}\right)\,\difd s.
\end{equation}
\end{subequations}

Although the integrals in \rf{eq:etameanvar} must, in general, be computed numerically,
they can be evaluated asymptotically at locations deep into the medium.
When $\La \ll x \ll \Lab$, rescaling the integration variable as $ s = \sqrt{x} s^{\prime} $, we see that
\begin{eqnarray}
&&\Eta(x,\sqrt{x} s^{\prime}) =
\tanh \left( 2\tau (a\sqrt{x}s^{\prime}+bx) +
\ln\frac{\left\vert d_+\right\vert}{\left\vert
d_-\right\vert}\right) \nonumber \\ && \sim \sgn s'
\label{eq:etaapprox}
\end{eqnarray}
while the Gaussian integration factor becomes independent of $ x $.  
We can therefore evaluate the $ \La \ll x \ll \Lab $ asymptotics of both the mean and variance by the asymptotic replacement (\ref{eq:etaapprox}), and find that
$ \langle \sin 2 \eta (x) \rangle \sim 0 $ and
$ \sigma^2_{\sin 2\eta} (x) \sim 1 $
 for $ \La \ll x \ll \Lab $.

On the other hand, if $ x \gg \Lb, \Lab $, we have under the same rescaling
\begin{eqnarray}
&&\Eta(x,\sqrt{x} s^{\prime}) =
\tanh \left( 2\tau (a\sqrt{x}s^{\prime}+bx) +
\ln\frac{\left\vert d_+\right\vert}{\left\vert
d_-\right\vert}\right) \nonumber \\ && \sim - \sgn b,
\end{eqnarray}
and so
we find that $\langle \sin  2 \eta  \rangle \sim - \sgn b$ and $\sigma^2_{\sin  2 \eta} \sim 0$ for large $x$.

To recapitulate, we have that for large $ x $
\begin{subequations} \label{eq:etaasy}
\begin{align}
  \langle \sin 2 \eta \rangle &\sim \begin{cases} - \sgn
 b,   & x \gg \Lb, \Lab,\\
 0, &   \La \ll x \ll \Lab,\end{cases}\\[12pt]
 \sigma^2_{\sin  2 \eta} &\sim \begin{cases} 0,  \hspace*{1cm} & x \gg \Lb, \Lab,\\
 1, &  \La \ll x \ll \Lab.\end{cases}
\end{align}
\end{subequations}
Again, we must recall from Eqs.~\rf{eq:lb}, \rf{eq:la}, and~\rf{eq:lfluc} that the case $x \gg \Lb, \Lab$ corresponds to only nonzero average initial population density difference $b=\langle\alpha(x)\rangle$, and the case $\La \ll x \ll \Lab$ also contains the case $b=0$.
We therefore see that all solitons with nonzero bias $b=\langle\alpha(x)\rangle$ will, with probability one,
collapse into a permanent circular polarization.
When  $b=0$, the large $x$ results for the mean and variance of $\sin 2 \eta$ in Eqs.~\rf{eq:etaasy} show that $ \langle [\sin 2 \eta (x)]^{2}  \rangle
= \sigma^2_{\sin  2 \eta}(x) + \langle\sin \eta (x) \rangle^{2} \rightarrow 1 $, which along with $ \langle \sin 2 \eta \rangle \rightarrow 0 $ and the inequality $ |\sin 2 \eta (x)|^{2} \leq 1 $, implies  that $ \sin 2 \eta (x) $ must converge at large $ x $ to
a random variable concentrated at $ \pm 1 $, with equal weight. In other words, for large $x$, the light polarization becomes one of the two circular polarizations with equal probability.

\paragraph{Probability Distribution of the Angle of Ellipticity}
\label{par:pdfpolar}
The large $x$ asymptotic behavior of the ellipticity angle $\eta$, described at the end of the previous section, can also be seen from developing the exact formula for the  cumulative distribution function $F_{\eta}(s;x)$ of $\eta$, which can be computed from the expression for $\eta$ in Eq. \rf{eta1} via a formula analogous to Eq.~\rf{cumuleq}.  In this way, we find
\begin{eqnarray}
&&F_{\eta}(s;x)=\Prob\left\{W(x) \geq \frac{1}{2a\tau}\biggl(\tanh^{-1}(\sin 2s ) \right.\nonumber\\ && \left.- 2\tau bx -\ln\frac{|d_+|}{|d_-|}\biggr)\right\}.
\end{eqnarray}
Since $W(x)$ is normally distributed with mean 0 and variance $x$, we compute
\begin{eqnarray}\label{etacdf}
&&F_{\eta}(s;x)=\frac{1}{2}\left\{1-\chi\left[x,\frac{1}{2a\tau}\biggl(\tanh^{-1}(\sin 2s )\right.\right. \nonumber\\ &&\left.\left. - 2\tau bx -\ln\frac{|d_+|}{|d_-|}\biggr)\right]\right\},
\end{eqnarray}
where $\chi(x,u) = \erf\left(u/\sqrt{2x}\right)$, with the function $\erf(\cdot)$ defined in Eq.~(\ref{error}).
After differentiating Eq.~\rf{etacdf} with respect to the parameter $s$, we derive the probability density function
\begin{eqnarray}\label{eq:petaxs}
&& p_{\eta}(s;x)=\frac{1}{\sqrt{2\pi x}\, a|\tau|\cos 2s}\nonumber \\ &&\times \exp\left\{-\frac{\Bigl[\tanh^{-1}(\sin 2s)-2\tau bx-\ln|d_+|/|d_-|\Bigr]^{2}}{8a^{2}\tau^{2}x}\right\}.\nonumber\\&&
\end{eqnarray}
Note that formulae \rf{etacdf} and \rf{eq:petaxs} are only valid for $-\pi/4\leq s\leq \pi/4$, the range in which the ellipticity angle $\eta$ is defined.

\begin{figure}
\includegraphics[scale=.4]{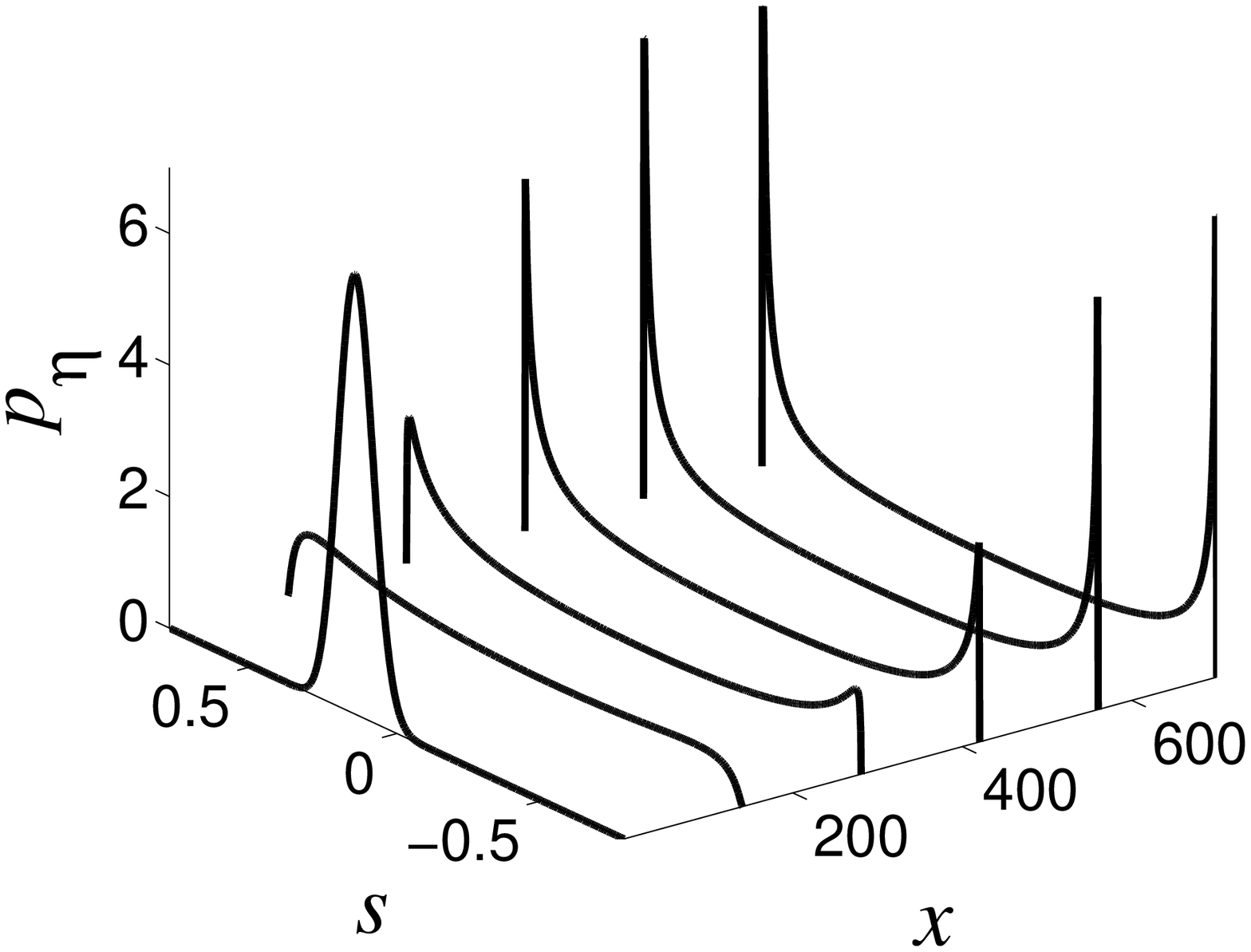}
\includegraphics[scale=.4]{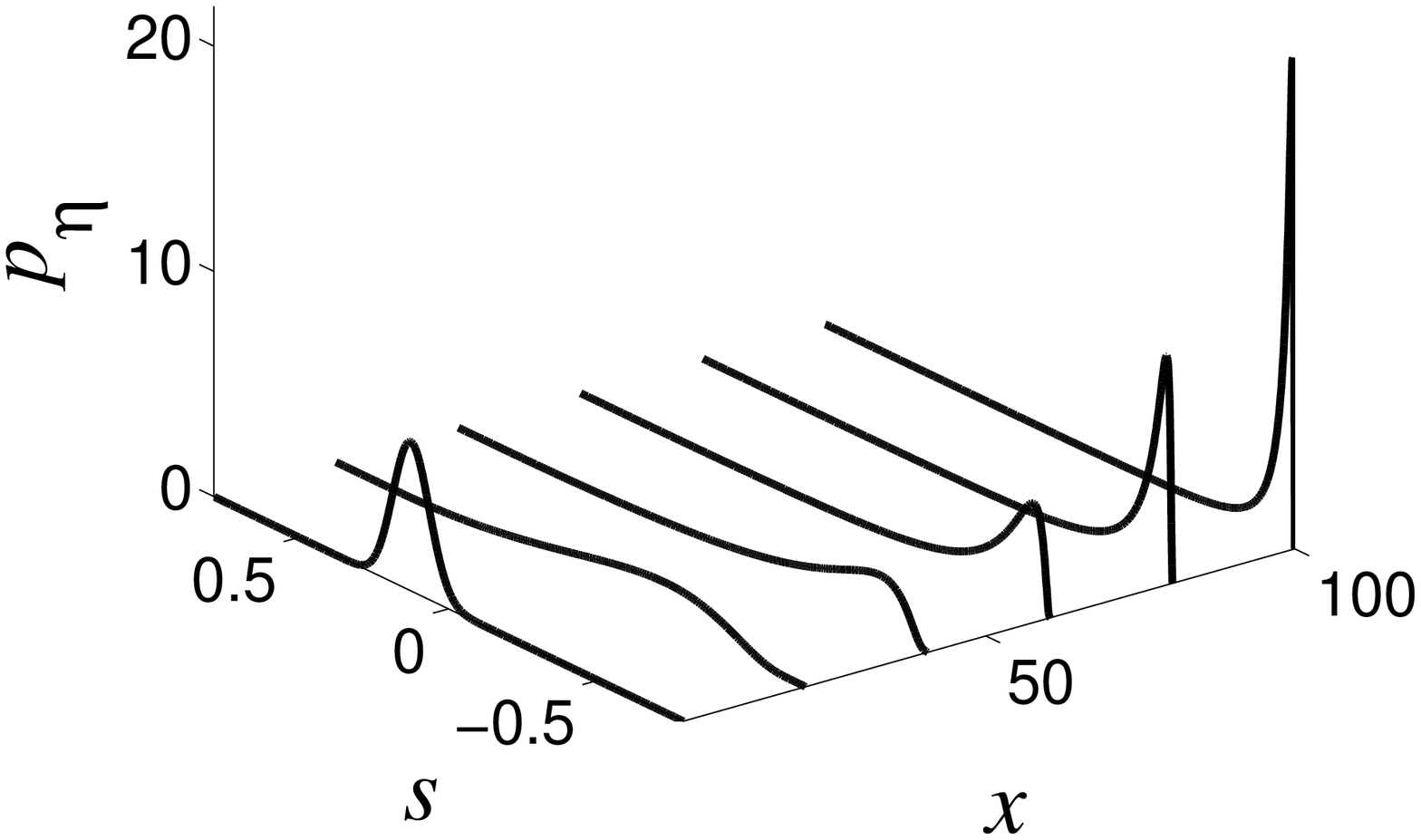}
\caption{Probability density function, $p_\eta(s;x)$, with $\beta=1$, $\gamma=1$,  $\varepsilon=0.1$, $d_+=i$, $d_-=3i/4$,  $a=1$.  Top: no bias, $b=0$.  Bottom: nonzero bias, $b=0.3$. }
\label{fig:peta}
\end{figure}

Very far into the active medium, that is, for large values of $x$, the distribution function $p_\eta(s;x)$ in \rf{eq:petaxs} clearly attains very small values at all $s$ away from $s=\pm\pi/4$, while at $s=\pm\pi/4$, it exhibits singularities.  Indeed, as remarked in Section~\ref{sec:meaneta}, the large $ x $ asymptotics of the mean and variance of the angle of ellipticity $ \eta (x) $ imply that its probability distribution must concentrate at one or both of the values $ s = \pm \pi/4 $ corresponding to the two circular polarizations:
\begin{equation}\label{eq:petalong}
p_{\eta}(s;x)\sim\begin{cases}
\displaystyle \frac{1}{2}\left[\delta\left(s-\frac{\pi}{4}\right)+\delta\left(s+\frac{\pi}{4}\right)\right], & b=0,\\[12pt]
\displaystyle\delta\left(s+\frac{\pi}{4}\right), &b>0,\\[12pt]
\displaystyle\delta\left(s-\frac{\pi}{4}\right),&b<0,
\end{cases}
\end{equation}
where
$\delta(\cdot)$ is the Dirac delta function.

The discussion in the preceding paragraph shows that for very
large distances $x$ into the medium, the soliton will mostly be confined to one of the two circular
polarizations.   For nonzero average initial population density difference, $\langle\alpha (x) \rangle=b\neq 0$, this polarization is fixed by the sign of $\langle\alpha (x)\rangle=b$ and with probability one eventually stops switching.   For $\langle\alpha (x)\rangle=b=0$,
at large distances into the medium, the
soliton stays in one of the two circular polarizations for most
of the time, switching intermittently between them.  The dynamics of the switching will be discussed in Section~\ref{sec:dynpol}.

\section{Dynamics of Polarization Switching}
\label{sec:dynpol}
Having developed explicit formulas for the soliton statistics as functions of depth into the optical medium when the initial population difference $\alpha(x)$ is random, we now provide a brief quantitative description of the dynamics of polarization switching.   We begin in Section~\ref{sec:dynlength} by identifying some key length scales to describe the essential features of the polarization dynamics.  Then, in Section~\ref{sec:wiendyn}, we present some analytical results for the polarization switching dynamics in the  Wiener process approximation.

\subsection{Length Scales of Polarization Switching Dynamics}
\label{sec:dynlength}

Because the asymptotic states of light pulses interacting with a $\Lambda$-configuration medium are given by the two circular polarizations, crossing the linear polarization represents a key reference point on the pulse trajectories.   In particular, the key elementary stages of the polarization switching can be cast in terms of this crossing:  The pulse will generally evolve from a linear (or elliptical) polarization to a nearly circular polarization, and then possibly eventually return to a linear polarization, from which it could return to its previous or the opposite circular polarization.  The characteristic length scales corresponding to the transitions between a linear and circular polarization and the successive returns to a linear polarization  need not be the same, because the pulse may reside near a circular polarization for long distances before returning to a linear polarization.   Moreover, for polarization switching in the presence of a nonzero bias, $b=\langle \alpha (x)\rangle \neq 0$, a pulse will eventually remain in one circular polarization forever, and so it is important to introduce a characteristic distance after which there is no more switching.

In view of the discussion in the previous paragraph, we can identify three distinct lengths of interest associated with the light pulse polarization switching process:  the distance between successive switches, the length scale over which
the switching process manifests itself when it does occur, and the distance into the medium over which switching continues.  One can distinguish these length scales more precisely by defining the
following \emph{random} distances:
the \emph{switching transition distance} $ \Xtra $ over which the light pulse polarization evolves from a linear state ($ \eta = 0 $) to a nearly circular polarization state of \emph{either} orientation  ($|\eta| = \pi/4 - \eta_c $), the \emph{interswitch distance} $ \Xint$  over which the light pulse polarization evolves from a linear state ($ \eta  = 0 $) to a nearly circularly polarized state ($|\eta| > \pi/4 - \eta_c$ for some fixed $ \eta_c  >0 $) of either orientation and back to a linear state ($\eta = 0 $),
and
the \emph{switching region depth} $ \Xdepth $ beyond which the light pulse polarization remains forever  in one of the circularly polarized states  $ |\eta| > \pi/4 - \eta_c $ for all greater distances.
As noted above,
the interswitch distance $ \Xint $ is not necessarily the same order of magnitude of the switching transition distance $ \Xtra$ because $ \Xint $ also includes the distance over which the soliton remains in a circular polarization before returning to a linearly polarized state.

\subsection{Polarization Dynamics in Wiener Process Approximation}
\label{sec:wiendyn}

From the equation~\rf{eta1} describing the spatial dependence of the ellipticity angle $\eta$ on the position $x$ along the medium sample, one can see that the distances $ \Xtra $, $ \Xint$, and $ \Xdepth $ depend solely on the level-crossing properties of the Wiener process $2\tau A(x) + \ln (|d_+|/|d_-|)$.   In particular, computing $ \Xtra $, $ \Xint$, and $ \Xdepth $ is equivalent to finding the positions along the medium sample for which this process first reaches the absolute value $\tanh^{-1}\left[\cos(2\eta_c)\right]$ after having passed through the origin, first returns to the origin after such an excursion, and remains further from the origin than this absolute value for all subsequent  $x$.
We consider separately the case in which the initial population density difference $\alpha(x)$ in the medium has no bias ($ b= \langle \alpha (x) \rangle = 0 $) and in which it does have bias ($ b = \langle \alpha (x) \rangle \neq 0 $).

\subsubsection{Case of no Medium Bias}
\label{sbsb:nobiaspol}
When the initial population-density bias vanishes, Eq.~(\ref{eq:etaasy}) implies that the polarization observed at any given position deep into the medium is likely to be circular, with probability $1/2 $ for each orientation.  From a dynamical perspective, in fact the polarization switches infinitely often, arbitrarily far into the medium, with probability one, because the Wiener process is recurrent in one dimension.  Consequently $ \Xdepth = \infty $ 
with probability one.
The polarization does indeed reside over great distances within one or the other circular polarization state, punctuated occasionally (but persistently) by switches (over relatively short distances) into the opposite polarization.

\begin{figure}[h]\begin{center}
\includegraphics[scale=.45]{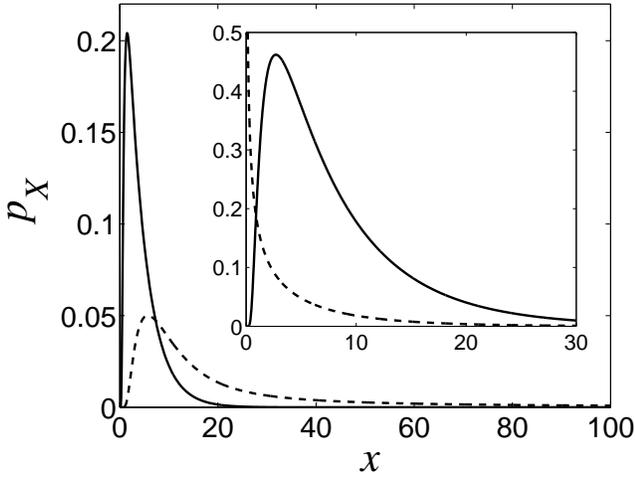}
\caption{\label{fig:pxs}The distributions $\pXtra (x)$ (solid line) and $\pXint (x) $ (dashed line) for a medium with no bias, $b=\langle \alpha(x)\rangle = 0$, computed using Eqs.~(\ref{eq:pxs}) with $\lswtrans=4.53$.   Note the  faster decay of the distribution $\pXtra (x)$.   Inset: The distributions $\pXtra (x)$ (solid line) and $p_{\Xstay} (x) $ (dashed line) for a medium with strong bias, computed using Eqs.~(\ref{eq:pxtrab}) and~(\ref{eq:pxstay}), respectively, with $a=1$, $b=-0.5$, and $\lswtrans=4.26$.}\end{center}
\end{figure}

Finding the statistics of the distances $ \Xtra $ and $ \Xint$ is equivalent to finding the corresponding distance statistics for the Wiener process $2\tau a W(x)$ without drift.   
Therefore, we can apply the well-known first-passage-time formulas (see, for example, Sec. 7.3 in~\cite{karlin75} or Sec. 2.8 in~\cite{karatzas91}), including their convolution computed via Laplace transforms, as described in Appendix~\ref{sec:conv},
to 
obtain the following expressions for the probability density function $ \pXtra (x)$ for the transition switching distance $ \Xtra $, and the probability density function $ \pXint (x)$ for the interswitch distance $ \Xint $, using the definitions in Sec.~\ref{sec:dynlength}:
\begin{subequations}\label{eq:pxs}
\begin{align}
\pXtra (x) &= \sqrt{\frac{2\lswtrans}{\pi x^3}}  \notag \\ &\times \sum_{n=-\infty}^{\infty}
(4n+ 1) \exp \left[- (4n+1)^2 \frac{\lswtrans}{2x}\right], \label{eq:pxtra}
\end{align}
\begin{align}
\pXint (x) &= \sqrt{\frac{2\lswtrans}{\pi x^3}}  \notag \\ &\times
\left\{ \sum_{n=0}^{\infty}
(4n+ 2) \exp \left[- (4n+2)^2 \frac{\lswtrans}{2x} \right] \right. \notag \\
& \left. \qquad -  \sum_{n=1}^{\infty}
4n \exp \left[- 16 n^2 \frac{\lswtrans}{2x}\right]\right\} , \label{eq:pxint}
\end{align}
\end{subequations}
where
\begin{equation}
 \lswtrans = \lswint = \left[\frac{1}{2a |\tau|}\tanh^{-1} \left(\cos (2 \eta_c)\right)\right]^2
 \label{eq:nobiaslength}
 \end{equation}
set the corresponding switching length scales (cf. the distance $L_a$ in Eq.~\rf{eq:la}.)  These two distributions are depicted in Fig.~\ref{fig:pxs}.

\begin{figure}
\begin{center}
\includegraphics[scale=.4]{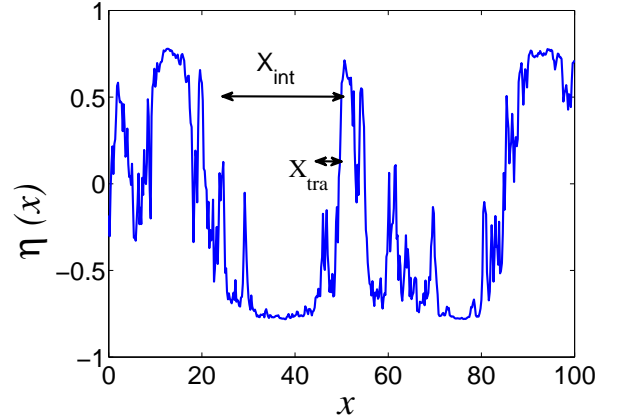}
\end{center}
\caption{Polarization state in a medium without bias under the Wiener Process Approximation ($a=1.00 $, $ b=0$).  The length scales for the transitions and distances between switches are here $ \lswtrans = \lswint = 4.53 $, but note that the distance between switches has higher probability to take values large compared to this typical length scale.\label{fig:nobias}}
\end{figure}

Note that the interswitch length and transition switching length appearing in the probability density functions are the same, but the probability distributions are quite different.  In particular, the transition switching distance has finite mean and variance:
\begin{equation}\label{eq:xtrameans}
\langle \Xtra \rangle = \lswtrans, \qquad
\sigma^2_{\Xtra} = \frac{2}{3} \lswtrans^2.
\end{equation}
On the other hand, the probability density function for the interswitch distance $ \Xint $ is so slowly decaying that $ \Xint $ has an infinite mean, even though we have identified a finite interswitch length scale~(\ref{eq:nobiaslength}).  The meaning of this is that while many polarization switching events do take place with interswitch distances comparable to $ \lswint $, on occasion a much longer distance is observed between polarization switches, and these rare events still have a large enough probability to imply an infinite mean interswitch distance.  The transition switching distances are however much more likely to be on the order of $ \lswtrans $, as can be seen from Eq.~\rf{eq:xtrameans}.  That is, the polarization will fairly often tarry in a circular polarization state for a distance much larger than $ \lswint $, but it will usually move out of the linear polarization state over the length scale $ \lswtrans $, as shown in Fig.~\ref{fig:nobias}.  This is naturally reflected in the polarization statistics developed in Eq.~\rf{eq:etaasy}, which indicate that the polarization will, deep in the medium, tend to be in one of the two circular polarization states.

Another quantifiable statistic for the case of no bias in the initial population density is the fraction $\Phi$ of the length over which the polarization $ \eta $ takes a certain sign.  
One can show (see~\cite{Feller} or Sec. 1.4.4 in~\cite{borodin02}) that 
its probability distribution is given by the arc sine law, which has the feature of having a rather large probability to take values near $ \phi = 0 $ or $ \phi = 1 $, meaning that even though the medium is unbiased, an individual realization is rather likely to spend most of its observed time in one or the other circular polarization state: 
\begin{align*}
\Prob (\Phi < \phi) &= \frac{2}{\pi} \sin^{-1} \sqrt{\phi} = \int_0^{\phi} p_{\Phi} (\phip) \, \difd \phip, \\
p_{\Phi} (\phi) &= \frac{1}{\pi \sqrt{\phi(1-\phi)}}. \nonumber
\end{align*}

\subsubsection{Case of Nonzero Medium Bias}
If  the initial population of the atomic ground levels in the medium does have a bias
($ b = \langle \alpha (x) \rangle \neq 0$), then polarization switches due to random fluctuations can occur for a while, but the bias in the population density of the optical medium will with probability one eventually
 collapse forever into the preferred circular polarization state associated with the sign of the initial population density bias $b= \langle \alpha (x) \rangle $~\cite{Byrne03}.
More precisely, if the bias $ b $ has opposite sign to
that of the ellipticity angle,
\begin{equation}
  \etainit = \oh \sin^{-1} \left[\tanh \left(\ln \frac{|d_+|}{|d_-|}\right)\right],  \label{eq:etainit}
  \end{equation}
  of the pulse upon entering the medium, 
the polarization of the soliton will proceed through the following stages: 
achieving a linear polarization for the first time after a distance $ \Xlinfirst $,
moving into a favored circular polarization state over a subsequent distance $ \Xtra $,
and
never leaving this ultimate circular polarization state after a subsequent distance $ \Xstay $.

The polarization switching depth $ \Xdepth $ would then be the sum of the lengths corresponding to these stages $ \Xdepth = \Xlinfirst + \Xtra + \Xstay $.  The interswitch distance $ \Xint $ is not well defined for the case of a medium with bias because eventually the soliton will stop switching.  If the bias were of the same sign as the initial polarization state, then $ \Xlinfirst $ would be meaningless and should just be treated as $ 0 $ in what follows, and the probability distribution for $ \Xtra $ would be changed only by replacing the length scale for $ \lswtrans $ in Eq.~\rf{eq:ltrabias} below by
\begin{equation*}
 \lswtrans =\frac{1}{2 |\tau  b|} \left[\tanh^{-1} (\cos (2 \eta_c)) - \left|\ln \frac{|d_+|}{|d_-|}\right|\right].
 \end{equation*}

First we develop formulas for the statistics of these lengths, then discuss the qualitative differences between soliton polarization evolution in a medium with weak bias and with strong bias.
The probability density function $ \plinfirst (x) $ for the distance $ \Xlinfirst $ until a linear polarization $ \eta = 0 $ is first reached can be expressed through another first passage time formula for Brownian motion with drift (see  Sec. 7.5 in~\cite{karlin75}),
\begin{displaymath}
\plinfirst (x) = \frac{|b| \lswfirst}{a \sqrt{2 \pi x^3}} \exp\left[- \frac{b^2 (\lswfirst - x)^2}{2 a^2 x}\right],
\end{displaymath}
where
\begin{equation}
 \lswfirst =\frac{1}{2 |\tau  b|} \ln \frac{|d_+|}{|d_-|}  \label{eq:biasdepth}
 \end{equation}
  and
\begin{subequations}
\begin{align}
 \langle \Xlinfirst \rangle& = \lswfirst, \\
\sigma^2_{\Xlinfirst} &=  \frac{a^2 \lswfirst}{ b^2}.
\end{align}
\end{subequations}

After reaching the linear polarization state, the polarization will tend to move toward its favored circular polarization state, reaching it after a further distance $ \Xtra $ which has probability density function
(see Sec. 7.5.5 in~\cite{karlin75})
\begin{equation}\label{eq:pxtrab}
\pXtra(x) = \frac{|b| \lswtrans}{a \sqrt{2 \pi x^3}} \exp\left[- \frac{b^2 (\lswtrans -  x)^2}{2 a^2 x}\right].
\end{equation}
where
\begin{equation}\label{eq:ltrabias}
 \lswtrans =\frac{1}{2 |\tau  b|} \tanh^{-1}[ \cos (2 \eta_c)]
 \end{equation}
and
\begin{subequations}
\begin{align}
 \langle \Xtra \rangle& = \lswtrans,\label{eq:xtra} \\ 
\sigma^2_{\Xtra} &= \frac{a^2 \lswtrans}{ b^2}. 
\end{align}
\end{subequations}

Once the polarization achieves a value $ |\eta| > \pi/4  - \eta_c $  corresponding to a circular polarization with the same sign as $ b $, it will continue to revisit the linear polarization state $ \eta = 0 $ over a further distance $ \Xstay $, for which a \emph{last} passage time formula (see Sec. IV.5 in~\cite{borodin02}) shows that it is distributed as  $  \Lfluc G^2$,
where $ G $
is a standard Gaussian random variable with mean zero and unit variance, and $\Lfluc \equiv a^2/b^2$, as defined in Eq.~\rf{eq:lfluc}, sets the length scale over which switching behavior continues.  In other words, $ \Xstay $ is governed by a  $\chi^2$ distribution with one degree of freedom, and has probability density~\cite{Feller}
\begin{equation}
p_{\Xstay}(x)
=\sqrt{\frac{\Lfluc}{2 \pi x}} \exp\left(-\frac{x}{\Lfluc}\right), \label{eq:pxstay}
\end{equation}
and mean $ \langle \Xstay \rangle =\Lfluc$.
The distributions $\pXtra(x)$ in Eq.~\rf{eq:pxtrab} and $p_{\Xstay}(x)$ are depicted in the inset of Fig.~\ref{fig:pxs}.

The qualitative character of the soliton trajectory will depend on the ratio
\begin{equation}
 \frac{\langle \Xstay \rangle}{\langle \Xtra \rangle}, \label{eq:biasratio}
\end{equation}
where, from (\ref{eq:pxstay}), $ \langle \Xstay \rangle = \Lfluc= a^2/b^2 $, and $ \langle \Xtra \rangle = \lswtrans $ is given in (\ref{eq:xtra}).
The numerator  describes the distance over which the polarization continues to fluctuate between the two circular polarizations, whereas the denominator characterizes a single transition from linear to circular polarization.  From the formulas (\ref{eq:xtra}) and~(\ref{eq:pxstay}), we see that the numerator is a more sensitive function of the bias than the denominator, in particular diverging faster as $ b \rightarrow 0 $.
We consequently divide our subsequent discussion into two cases:
 the \emph{strong bias} regime in which $ \langle \Xtra \rangle \gg \langle \Xstay \rangle $,
and the \emph{weak bias} regime in which $ \langle \Xtra \rangle \ll \langle \Xstay \rangle $.
We always assume $ \langle \Xlinfirst \rangle \lesssim \langle \Xtra \rangle $ in what follows, which can be insured by simply taking a sufficiently small choice of $ \eta_c $ in the definition of circular polarization at the beginning of Subsection~\ref{sec:dynlength}.

\begin{figure}
\includegraphics[width=0.49\textwidth]{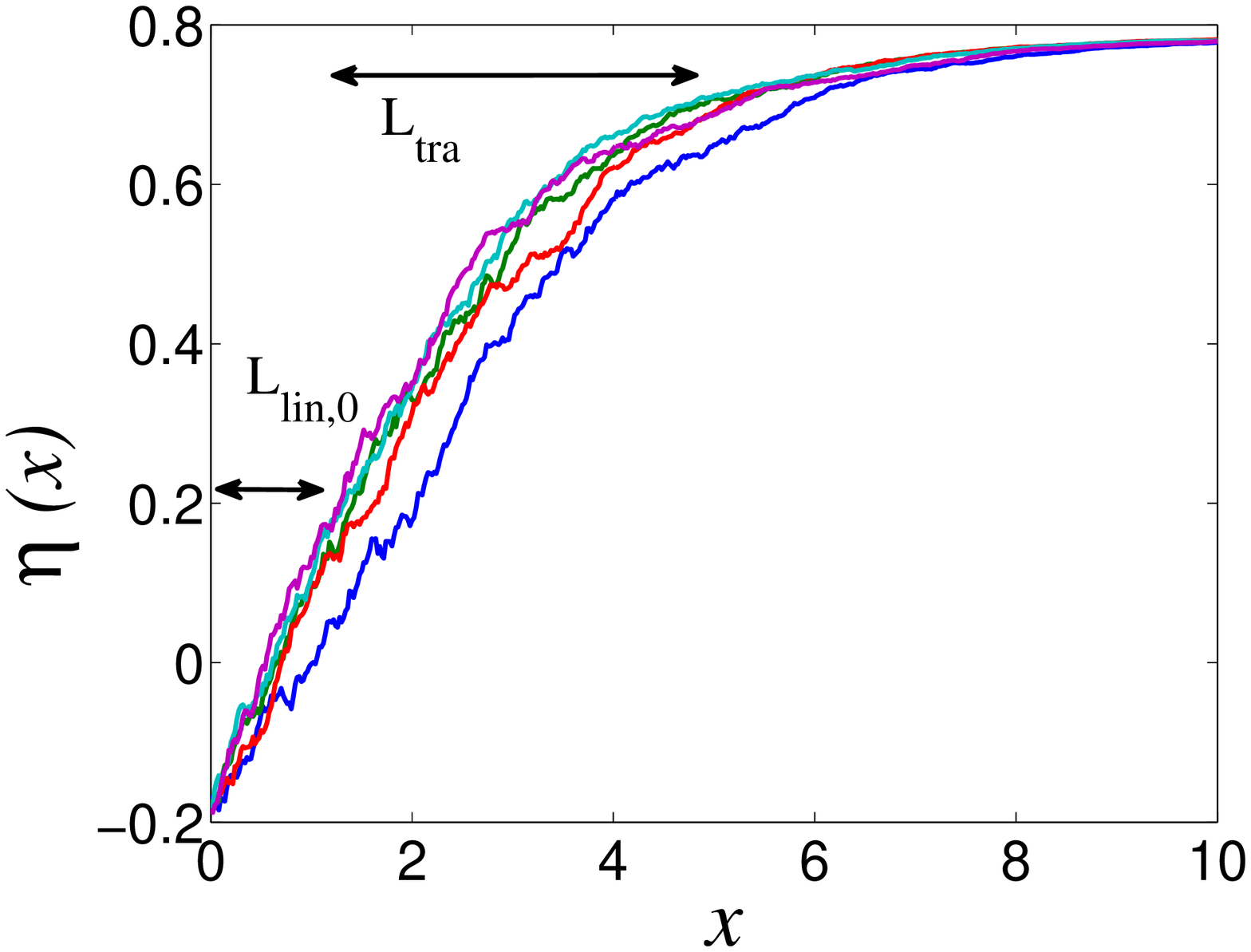}
\includegraphics[width=0.49\textwidth]{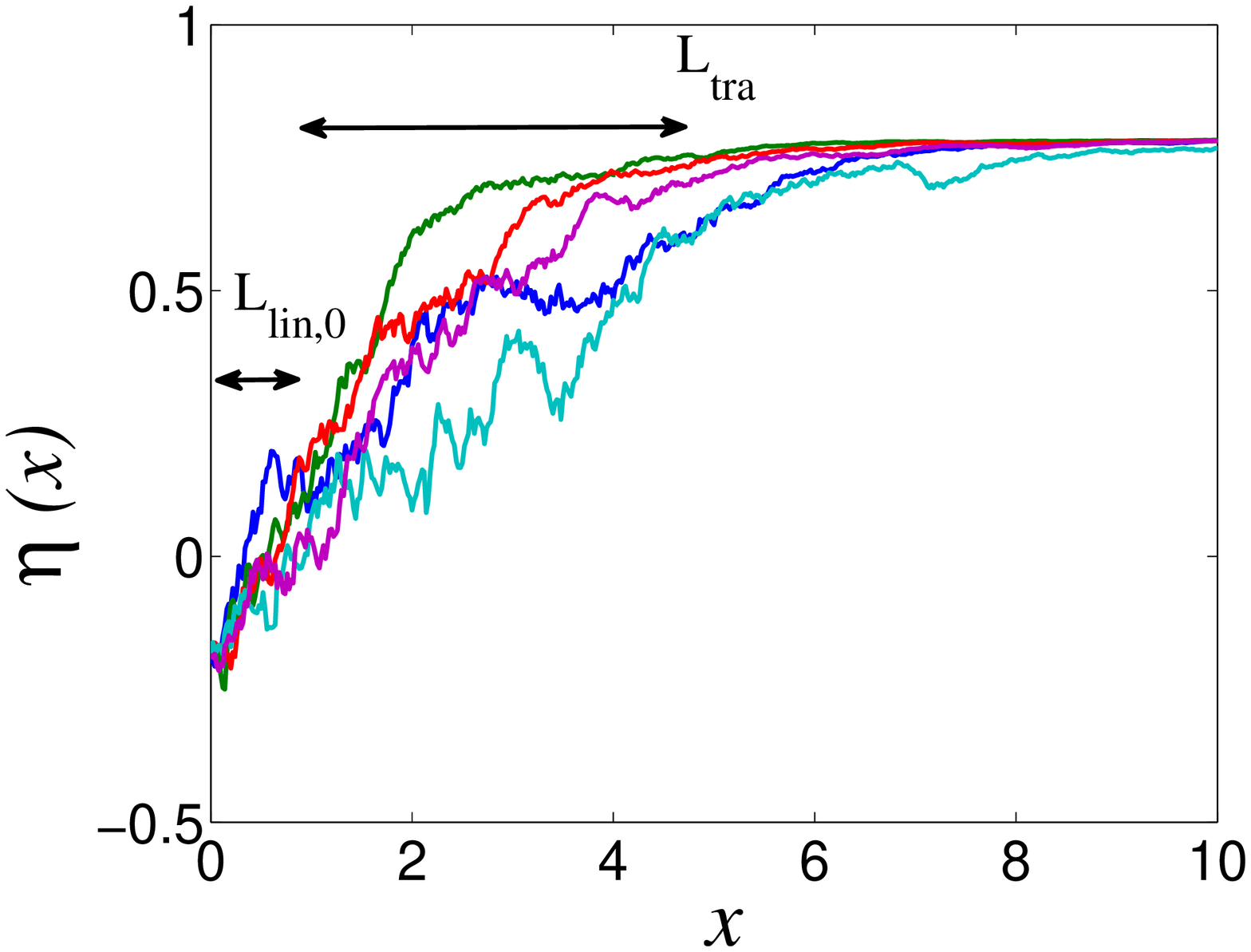}
\caption{Realizations of the soliton polarization for a medium with strong bias ($a=0.1$, $b=-0.5$,
$\lswfirst = 0.64$, $\lswtrans= 4.26 $, $ \lfluc = 0.04 $) on the top and a medium with not so strong bias
($ a=0.3$, $ b= -0.5 $, $ \lswfirst = 0.64 $, $ \lswtrans = 4.26$, $\lfluc = 0.36 $) on the bottom.\label{fig:strongbias}}
\end{figure}

\paragraph{Strong Bias Regime}
When $ \langle \Xtra \rangle \gg \langle \Xstay \rangle $, then the bias dominates the dynamics to the extent that the key distances characterizing the polarization dynamics are well-described by deterministic expressions. We will for the most part consider the typical case in this regime, in which
$ \langle \Xlinfirst \rangle \gg \langle \Xstay \rangle $ as well, and comment on what happens when this is not true later.  Proceeding under the assumption that $ \langle \Xlinfirst \rangle \gg \langle \Xstay \rangle $, the standard deviation of the distances $ \Xlinfirst $ and $ \Xtra $ is comparable to or smaller than their mean, so that these distances are indeed comparable to the deterministic length scales $ \lswfirst $ and $ \lswtrans $ with high probability.  Moreover, the distance $ \Xstay $ is negligible relative to these other distances. Consequently, the switching region depth is approximately deterministic with $ \Xdepth \sim \lswfirst + \lswtrans $.  This means that, with high probability, the soliton experiences exactly one polarization switch (as it moves into the favored polarization state), and does so in an approximately deterministic manner after a distance $ \lswfirst + \lswtrans $, as shown in the top panel of Fig.~\ref{fig:strongbias}.  When the ratio in Eq.~(\ref{eq:biasratio}) becomes order unity, the polarization switch becomes a bit more random (bottom panel of Fig.~\ref{fig:strongbias}), but usually multiple switches are not seen.

The case in which $ \langle \Xlinfirst \rangle \lesssim \langle \Xstay \rangle $ is only a minor modification of the above description, in that now $ \Xlinfirst $ behaves randomly but plays a negligible role in the dynamics since necessarily $ \langle \Xlinfirst \rangle \ll \langle \Xtra \rangle $ given the definition of the strong bias regime.

\paragraph{Weak Bias Regime}
For weak bias, when $ \langle \Xstay \rangle \gg \langle \Xtra \rangle $, the distances $ \Xlinfirst $ and $ \Xtra $ become highly variable.  This is to be expected since the limit of no bias $ b =0 $ involves a qualitatively different scenario described in Sec.~\ref{sbsb:nobiaspol}.  The weak bias regime  involves features of both the no bias and strong bias regimes.  On the one hand, the polarization will eventually collapse into a permanent circular polarization with the same sign as the bias $ b $.
 However, for the case of weak bias we expect several or even many visits to the linear polarization state before the ultimate collapse into a circular polarization state.  Indeed, in the limit of no bias, the linear polarization state is visited infinitely often, as discussed in Sec.~\ref{sbsb:nobiaspol}).

\begin{figure}
\includegraphics[width=0.43\textwidth]{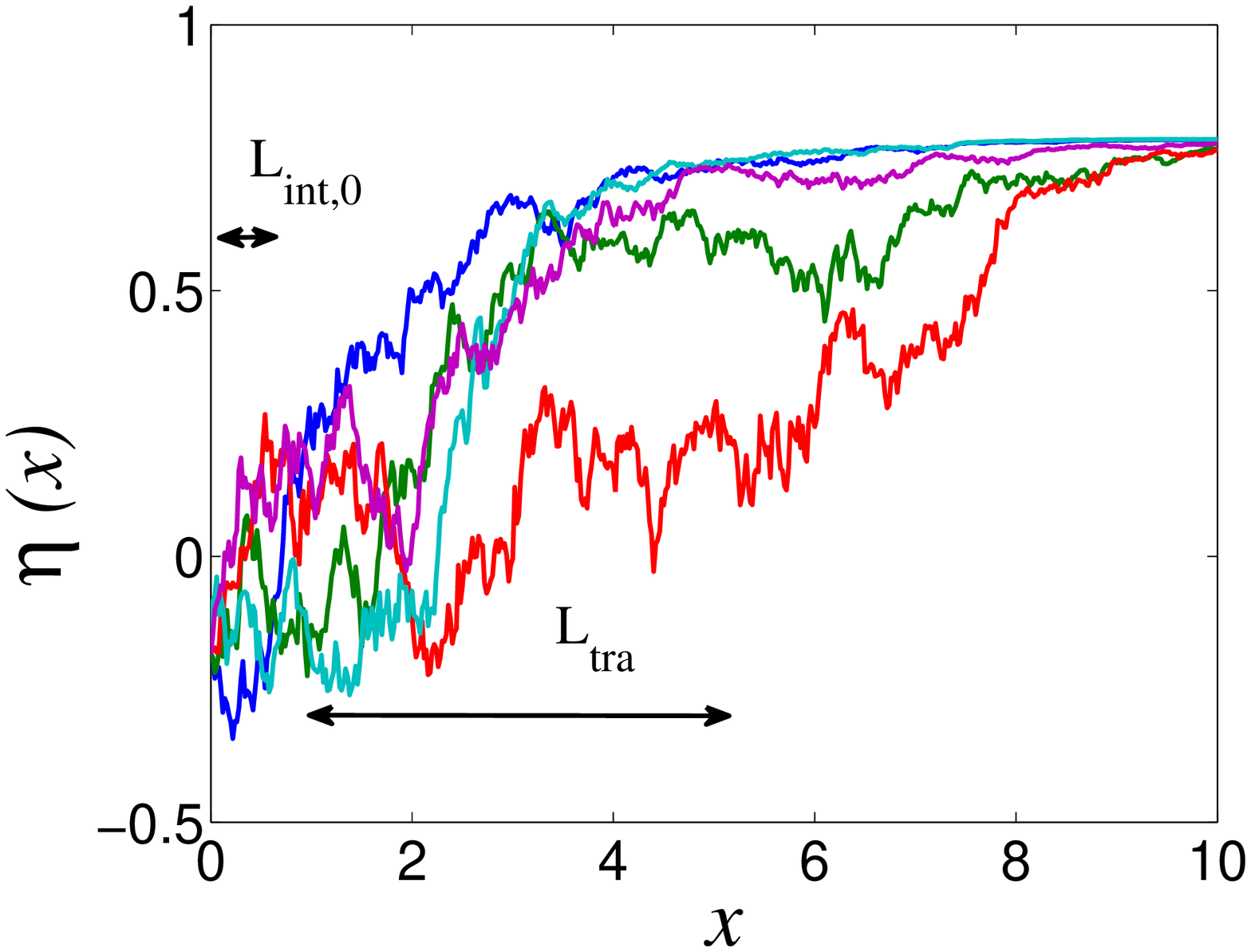}
\includegraphics[width=0.43\textwidth]{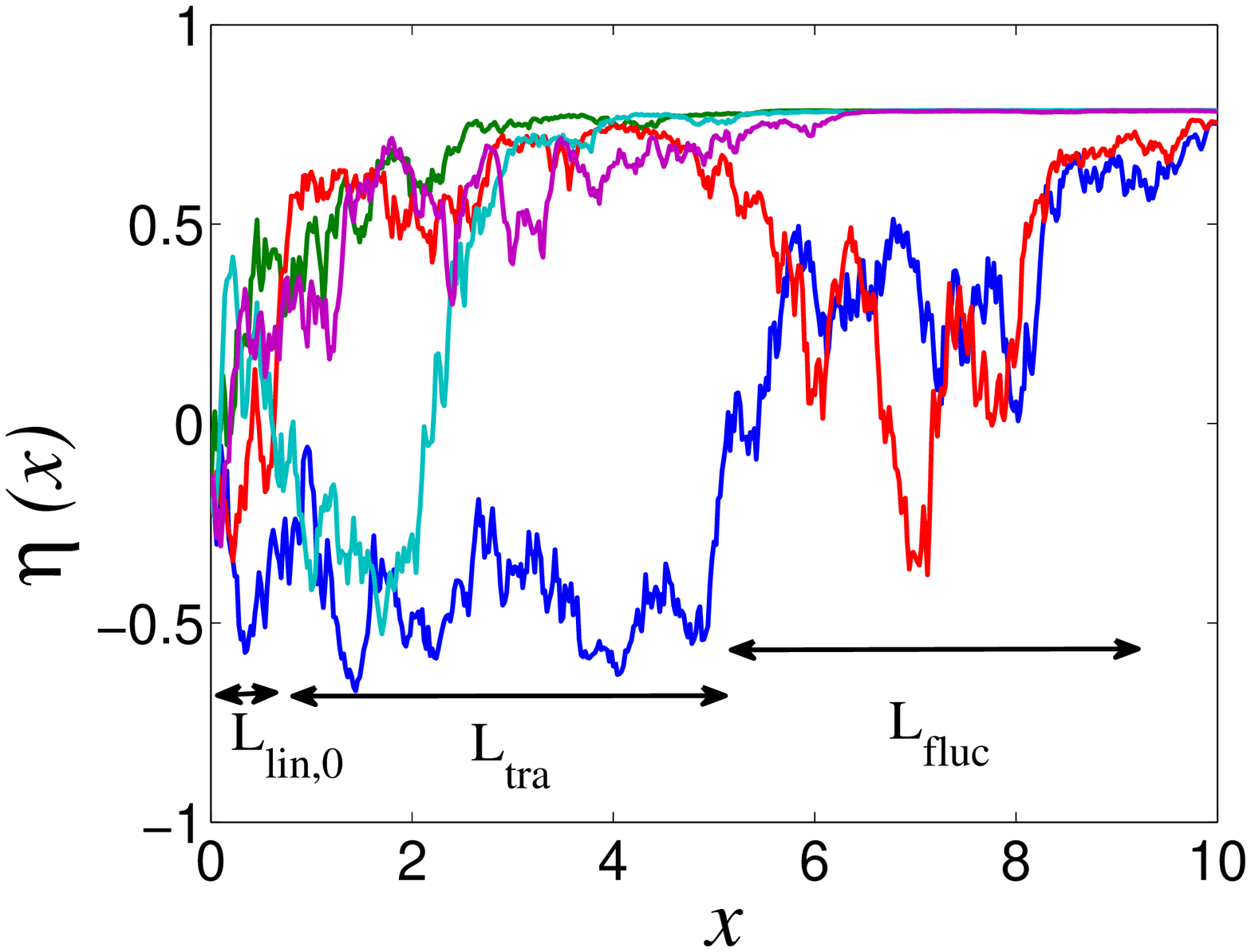}
\includegraphics[width=0.43\textwidth]{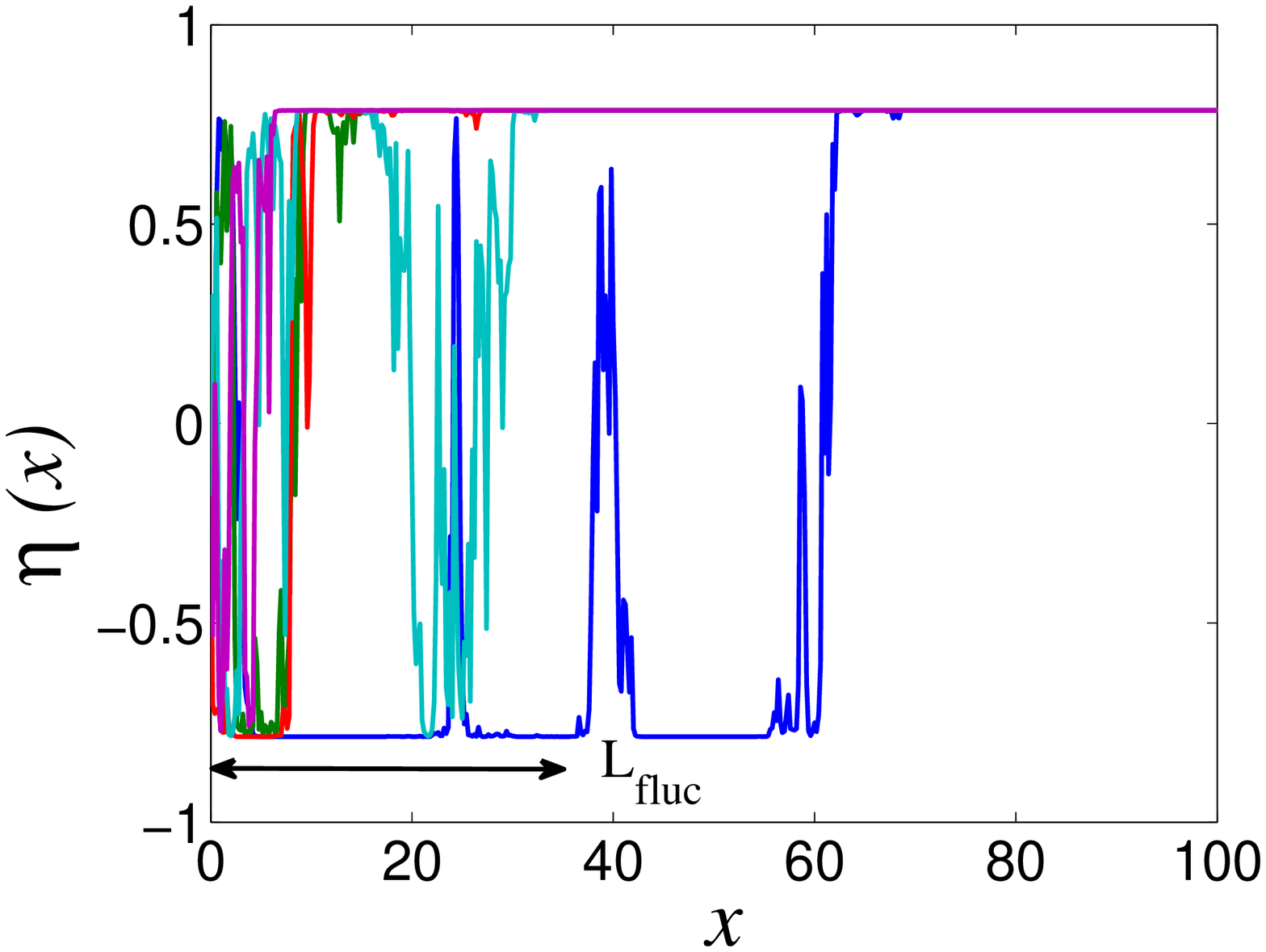}
\caption{Realizations of the soliton polarization for a medium with bias becoming progressively weaker from the top to bottom panel.  Top:  $a=0.5$, $b=-0.5$,
$\lswfirst = 0.64$, $\lswtrans= 4.26 $, $ \lfluc = 0.04 $; middle:
$ a=1.0$, $ b= -0.5 $, $ \lswfirst = 0.64 $, $ \lswtrans = 4.26$, $\lfluc = 4.00 $;
bottom:  $ a=3.0$, $b=-0.5$, $ \lswfirst = 0.64$, $\lswtrans = 4.26$, $\lfluc = 36.0$.}
\label{fig:weakbias}
\end{figure}

The dynamics are in fact dominated by the extended period of random polarization switching since the initial approach toward the favored circular polarization state is short by comparison
($ \langle \Xstay \rangle \gg \langle \Xlinfirst \rangle, \langle \Xtra \rangle $), though as noted above this initial approach has a highly random character because the standard deviations of $ \Xlinfirst $ and $ \Xtra $ are large compared to their means.  The polarization switching depth is therefore determined predominantly by the length scale $ \Xdepth \sim \Xstay $ which has statistics described in Eq.~(\ref{eq:pxstay}).
Figure~\ref{fig:weakbias} illustrates how, as the bias is weakened (proceeding downward),
the pulse polarization  switches randomly over an extended random distance $ \Xstay $ before finally collapsing into the favored polarization state.

\section{Conclusions\label{sec:conclusions}}
We have analyzed the resonant interaction of single-soliton light
pulses with a $\Lambda$-configuration  degenerate optical medium in
the idealized integrable Maxwell-Bloch approximation.   This is an
example of a phenomenon for which integrability and structural
disorder produce nontrivial stochastic nonlinear dynamics, yet whose
statistics can be analyzed in closed form.  We have found explicit
dependence of the soliton polarization on the average difference
between the initial populations of the degenerate lower sub-levels
along the medium sample, with infrequent but persistent random
switching between the two circular polarizations when this
difference vanishes, and almost certain asymptotic approach to one
of the two circular polarizations determined by this average
difference when it does not vanish.  Moreover, we have provided a precise quantification of the statistical dynamics of polarization switching, including probability distributions for the key distances describing transitions.

At least one question still remains about these results, which is whether
they are robust under the random fluctuations of the medium
polarization induced by finite-temperature effects.   Our preliminary numerical results
confirm that they should be robust.  These results, and also an analysis based on the full
evolution equations for the spectral data corresponding to random,
non-vanishing initial medium polarization variables, as presented
in~\cite{Byrne03}, will be described in a subsequent publication.

\begin{acknowledgments} We would like to thank David Cai,  Vladimir
Drachev, Andrei Maimistov, Katie Newhall, Valery Rupasov, David Shapiro,  Mikhail
Stepanov, and Eric Vanden-Eijnden for fruitful discussions. E.P.A. was supported by NSF and
DOE graduate fellowships. I.R.G. was partially supported by NSF grant
DMS-0509589, FTP S\&SPPIR, ARO-MURI award 50342-PH-MUR and State of
Arizona (Proposition 301), G.K. was partially supported by NSF grant
DMS-1009453.
\end{acknowledgments}

\appendix

\section{Correlation Length in the Optical Medium}
\label{sec:lcor}

In this appendix, we give a mathematically precise description of the correlation length, which is assumed to be effectively zero in the white noise approximation~\rf{eq:whitemodel} in Section~\ref{sec:wienercoarse}.  In general, without the white noise assumption (\ref{eq:corfun}), the correlation function of the population density difference $\alpha(x)$ in the medium is defined as
\begin{equation}\label{eq:corfun1}
\Ralph (x)=\langle [\alpha (\xp) - b] [\alpha (\xp+x)-b]
 \rangle ,
 \end{equation}
where we use the statistical spatial homogeneity of $ \alpha (x) $.  We further assume
\begin{equation}\label{eq:integr}
0 < \int_{0}^{\infty} \Ralph (x) \, \difd x < \infty,
\end{equation}
which means that  the correlations are sufficiently short-ranged.

Using Eqs. (\ref{eq:corfun1})  and (\ref{eq:integr}), the correlation length $\lcor$ is defined as
\begin{equation}\label{eq:lcor}
\lcor = \frac{1}{\Ralph (0)}\int_{0}^{\infty} \Ralph (x) \, \difd x= \frac{1}{\sigma^2_\alpha}\int_{0}^{\infty} \Ralph (x) \, \difd x.
\end{equation}
Physically, $ \lcor $ is the shortest length scale over which the intial population density difference $ \alpha (x) $ exhibits significant variations.

Under the
condition (\ref{eq:integr}) and for $x\gg \lcor$,
the functional central limit theorem for random fields~\cite{breiman92} implies that the function $ A(x)=\int_0^x\alpha(x)\,\difd x$, defined in \rf{Adef}, is statistically
equivalent at large $ x $ to the random process $ a W (x) + b x $,
where $b=\langle \alpha (x)\rangle$ is defined in (\ref{eq:mean}),
\begin{equation}
a = \left(2 \int_{0}^{\infty} \Ralph (x) \,
\difd x\right)^{1/2}, \label{eq:abdef}
\end{equation}
and $ W (x) $ is the standard Wiener process.

\section{Probability distributions for transition switching and interswitch distance in the absence of medium bias}
\label{sec:conv}

From the definition of $ \Xtra $, the formula (\ref{eta1}) for the evolution of the ellipticity angle, and the white noise approximation for the case of no medium bias ($ A(x) = a W (x) $), we deduce
\begin{align*}
\Xtra &= \min_{x \geq x_0} \left\{x - x_0: |\eta (x)| \geq \pi/4 - \eta_c \right\} \\
&= \min_{x \geq x_0} \left\{x - x_0 : \left|W(x) + \frac{1}{2 a \tau} \ln \frac{|d_+|}{|d_-|}\right| \geq \right. \\
& \qquad \qquad \qquad \qquad \left. \frac{1}{2 a |\tau|} \tanh^{-1} (\cos 2 \eta_c) \right\} \\
&= \min_{x \geq x_0} \left\{x \geq x_0 : \left|W(x) + \frac{1}{2 a \tau} \ln \frac{|d_+|}{|d_-|}\right| \geq 
\sqrt{\lswtrans} \right\}
\end{align*}
where in the above, $ x_0 $ is a position where $ \eta (x_0) = 0 $ (equivalently $ W(x_0) = - (1/2 a \tau) \ln (|d_+|/|d_-|) $).  
We see then that $ \Xtra $ is  just the distance of the first position $ x $ after $ x_0 $ at which the Wiener process $ W(x) $ escapes a given interval, given an initial position within that interval at $ x_0 $, also known as a first exit time.  The formula (\ref{eq:pxs}) then follows directly from the first exit time formula (2.8.24) in~\cite{karatzas91} and the translational invariance of the statistics of Wiener process increments.

The random variable $ \Xint $ has a two-step definition, which in mathematical terms can be translated as follows:
\begin{equation}
\Xint = \Xtra + \Xret \label{eq:xintsum}
\end{equation}
where the ``return distance" is defined as the distance over which the soliton returns from a nearly circular polarization state of either orientation ($ |\eta| = \pi/4 - \eta_c $) to a linear state ($ \eta = 0$):
\begin{align*}
\Xret 
&=  \min_{x \geq x_1} \{x - x_1: \eta(x) = 0 \} \\
& = 
 \min_{x \geq x_1} \left\{x - x_1 : W(x) = - \frac{1}{2 a \tau} \ln \frac{|d_+|}{|d_-|} \right\}  ,
 \end{align*}
with $ x_1 $ a position where $ \eta (x_1) = \pm  (\pi/4 - \eta_c) $ (equivalently $ W(x_1) = 
 -  (1/2 a \tau) \ln (|d_+|/|d_-|) \pm \sqrt{\lswtrans} $).
Because of the statistical reflection symmetry of the Wiener process, either sign of the $ \pm $ for the conditions at $ x_1 $ will give the same result, which is the first position $ x $ after $ x_1 $ at which the Wiener process $ W(x) $ achieves a certain value, given that it was situated at a different value at $ x _1 $.   This is known as a first passage time, and we can apply formula (2.8.5) in~\cite{karatzas91}, again using statistical translation invariance, to obtain the probability density function for $ \Xret $:
\begin{equation*}
\pXret (x) = \sqrt{\frac{\lswtrans}{2 \pi x^3}} \expe^{-\lswtrans/2x} \text{ for } x \geq 0.
\end{equation*}

Now, by the strong Markov property of the Wiener process, the summands $ \Xtra $ and $ \Xret $ in Eq.~(\ref{eq:xintsum}) are independent random variables, so the probability density of $ \Xint $ is the convolution of the probability densities of the summands.  Equivalently, the moment generating function
\begin{equation*}
M_{\Xint} (s) \equiv \langle \expe^{s \Xint} \rangle = \int_{0}^{\infty} \expe^{s x} \pXint (x) \, \difd x
\end{equation*}
is the product of the moment generating functions of the summands~\cite{karlin75}.  These moment generating functions can be computed through the Laplace transform identity:
\begin{equation}
\int_{0}^{\infty} \expe^{s t} \frac{b}{\sqrt{2 \pi t^3}} \expe^{-b^2/t} \, \difd t = \expe^{-b \sqrt{2 s}} \;\text{ for }\; b > 0,
\label{eq:laplace}
\end{equation}
which can be derived either by calculus tricks~\citep{friedmansde} or, more elegantly, by computing the probability density function  for the first passage time of Brownian motion by the reflection principle (Section 2.6 in~\cite{karatzas91}) and its moment generating function by the optional stopping theorem (Section 2.8 in~\cite{karatzas91}) and connecting these results.  Applying this Laplace transform identity, we obtain:
\begin{align*}
M_{\Xint} (s) &= M_{\Xtra} (s) M_{\Xret} (s) \\
& = 
2 \sum_{n=0}^{\infty} \expe^{-(4n+1) \sqrt{\lswtrans} \sqrt{2 s}} \expe^{-\sqrt{\lswtrans} \sqrt{2 s}} \\
& \qquad \qquad - 2 \sum_{n=-1}^{-\infty} \expe^{ (4n+1) \sqrt{\lswtrans} \sqrt{2 s}} \expe^{-\sqrt{\lswtrans} \sqrt{2 s}}  \\
&= 2 \sum_{n=0}^{\infty} \expe^{-(4n+2) \sqrt{\lswtrans} \sqrt{2 s}}
- 2 \sum_{n=-1}^{-\infty} \expe^{4n \sqrt{\lswtrans} \sqrt{2 s}}
\end{align*}
But then we can infer the probability density (\ref{eq:pxint})  for $ \Xint $ by again applying the Laplace transform identity (\ref{eq:laplace}) in reverse to each summand.

\end{document}